# Simple Fourier optics formalism for high angular resolution systems and nulling interferometry

**François Hénault**

*UMR 6525 CNRS H. Fizeau – UNS, OCA*

*Avenue Nicolas Copernic, 06130 Grasse – France*

In this paper are reviewed various designs of advanced, multi-aperture optical systems dedicated to high angular resolution imaging or to the detection of exo-planets by nulling interferometry. A simple Fourier optics formalism applicable to both imaging arrays and nulling interferometers is presented, allowing to derive their basic theoretical relationships as convolution or cross correlation products suitable for fast and accurate computation. Several unusual designs, such as a "super-resolving telescope" utilizing a mosaïcking observation procedure or a free-flying, axially recombined interferometer are examined, and their performance in terms of imaging and nulling capacity are assessed. In all considered cases, it is found that the limiting parameter is the diameter of the individual telescopes. The entire study is only valid in the frame of first-order geometrical optics and scalar diffraction theory. Furthermore, it is assumed that all entrance sub-apertures are optically conjugated with their associated exit pupils, a particularity inducing an instrumental behavior comparable with those of diffraction gratings.

*OCIS codes:* 070.0700; 110.2990; 110.5100; 110.6770; 350.1260.

# 1 Introduction

High angular resolution optical systems have been developed for more than one century, spanning from historical Michelson's interferometer [1] to the first fringes formed between two separated telescopes by Labeyrie [2]. Techniques of long baseline stellar interferometry are now widely accepted and understood [3], giving birth to modern observing facilities such as Keck interferometer, VLTI, or CHARA that are now intensively used to produce flows of high-quality scientific results, mainly in the field of stellar physics. In spite of this success some new ideas on ground or spaceborne multi-aperture observatories emerged in the two last decades, among which are infrared nulling interferometers dedicated to the search of extra-solar planets [4-6], and visible hypertelescopes having unsurpassed imaging capacities [7]. Much has already been written about the mathematical descriptions of these two different types of instruments (see for example refs. [8-10] and [11] respectively): it appears however that they can be regrouped together under a common and simple analytical formalism based on Fourier optics theory, also applicable to long baseline interferometry or to the tentative design of an hypothetical "super-resolving telescope". This formalism is briefly exposed in section 2 and a comprehensive roadmap to the various presented analytical relationships and numerical simulations is provided in section 3. The general properties of imaging arrays are discussed in section 4, as well as three examples of high angular resolution instruments. Some other important consequences applicable to nulling interferometry are derived in section 5. Planned future works and conclusions are finally presented in sections 6 and 7.

# 2 Formalism

In this section are described the main optical and geometrical characteristics of the considered multi-aperture, high angular resolution systems, and the analytical formalism relevant to their capacities for imaging and coupling into Single-Mode Waveguides (SMW).

## 2.1 Coordinates systems and scientific notations

The main reference frames employed on-sky, on the entrance and exit pupils, and in the image plane are represented in Figure 1 and Figure 2 and are defined as follows:

- The observed sky objects are located at infinite distance and are determined by a unitary vector $\mathbf{s_O}$ pointed along their direction. The cosine directors of $\mathbf{s_O}$ are approximately equal to (1,u,v), where u and v are the angular coordinates of the celestial target.

- The entrance pupil plane $X_PY_P$ is perpendicular to the main optical axis Z. It is assumed that all the sub-aperture lay in that plane, whose reference point is noted $O_P$.

- Similarly, all the output sub-pupils are arranged in a common exit pupil plane $X'_PY'_P$ referenced to point $O'_P$.

- The image plane X'Y' is attached to the focal point O' of the multi-aperture optical system. A unitary image vector $\mathbf{s}$ is associated to any point M' in the focal plane via the relation: $\mathbf{s} = \mathbf{O'_PM'} / |\mathbf{O'_PM'}|$.

It has to be underlined that all vectors appear in bold characters. However, in order to simplify certain analytical relationships presented in the remainder of the text, a condensed notation in bold italic characters has been adopted for vectors being perpendicular to the Z-axis, that will only be identified by their tip point (the origin point being either $O_P$, $O'_P$ or O'). For instance $\boldsymbol{P_n}$, $\boldsymbol{P'_n}$ and $\boldsymbol{M'}$ respectively stand for $\mathbf{O_PP_n}$, $\mathbf{O'_PP'_n}$ and $\mathbf{O'M'}$ in the whole paper.

## 2.2 General description of the optical system

Let us consider a multiple aperture, high angular resolution system such as depicted in Figure 3. It is basically composed of N individual collecting telescopes (noted L1_n on the Figure with index n related to the considered telescope, and $1 \leq n \leq N$), each having identical diameters D and focal lengths F. Between each telescope and the exit combiner L' are positioned identical optical trains (from lens L2_n to mirror M2_n. Note that all components represented by thin lenses could actually be reflective optics) in charge of collimating, compressing and conveying the optical beams. One basic assumption of this study is that each output sub-pupil is optically conjugated with its corresponding input sub-pupil: this implies that there must exist some divergent optics L2_n imaging each telescope entrance aperture (centred on point $P_n$ in plane $X_P Y_P$, see Figure 1 and Figure 3) on a corresponding area in the exit pupil plane (of fixed diameter D' and centred on point $P'_n$ in plane $X'_P Y'_P$). In practice the latter optical elements are commonly found in most multi-apertures interferometric facilities. Let finally $F_C$ be the focal length of the collimating lens L3_n, and assume that the interference fringes are formed and observed at the focal plane of a multi-axial beam combiner L', whose focal length is noted F'. The existing optical conjugations between the sky object and the image plane, on the one hand, and input and output pupils, on the other hand, imply that the pupil magnification ratio $m$ is equal to the ratio of the focal lengths of both collimating lens L3_n and entrance telescope L1_n:

$$m = D'/D = F_C/F, \qquad (1)$$

while the magnification $m_C$ of the combining optics writes:

$$m_C = F'/F_C. \qquad (2)$$

For a given celestial object of angular coordinates (u,v) pointed out by vector $\mathbf{s_O}$ and any point M' of coordinates (x',y') in the final image plane X'Y', the total Optical Path Difference (OPD) is equal to (see Figure 3):

$$\zeta_n = [P'_n H'_n] + [P_n H_n], \qquad (3)$$

for the $n^{th}$ interferometer arm, since the optical paths between points $P_n$ and $P'_n$ are constant throughout the whole Field of View (FoV) as a consequence of pupil conjugations. Using the condensed scientific notation described in section 2.1, Eq. (3) can be rewritten as a sum of scalar products very similar to the well-known diffraction gratings formula:

$$\zeta_n = \mathbf{O'_P P'_n}\ \mathbf{s'} + \mathbf{O_P P_n}\ \mathbf{s_O} = \boldsymbol{P'_n}\ \mathbf{s'} + \boldsymbol{P_n}\ \mathbf{s_O}. \tag{4}$$

In the frame of the first-order optics approximation in which this study will be entirely restricted, **s'** is approximated by:

$$\mathbf{s'} = \mathbf{O'_P M'} / \mid \mathbf{O'_P M'} \mid \approx (\mathbf{O'_P O'} + \mathbf{O'M'}) / F' \approx (\mathbf{O'_P O'} + \boldsymbol{M'}) / F'. \tag{5}$$

It is furthermore convenient to transpose the sky vector $\mathbf{s_O}$ into the X'Y' image plane where it corresponds to the vector $\mathbf{s'_O}$ and paraxial image $M'_O$ of Cartesian coordinates ($x'_O$, $y'_O$):

$$\mathbf{O'_P M'_O} = \mathbf{O'_P O'} + \mathbf{O'M'_O} = -m_C\ F\ \mathbf{s_O} = -F'\ \mathbf{s_O} / m \tag{6a}$$

and inversely: $$\mathbf{s_O} = -m\ (\mathbf{O'_P O'} + \mathbf{O'M'_O}) / F' = -m\ (\mathbf{O'_P O'} + \boldsymbol{M'_O}) / F'. \tag{6b}$$

Combining relationships (1), (2), (4), (5) and (6b) and noting that the scalar products $\boldsymbol{P'_n}\ \mathbf{O'_P O'}$ and $\boldsymbol{P_n}\ \mathbf{O'_P O'}$ are equal to zero finally leads to a condensed expression of the OPD $\zeta_n$ that is:

$$\zeta_n = \boldsymbol{P'_n}\, \boldsymbol{M'} / F' - m\, \boldsymbol{P_n}\, \boldsymbol{M'_O} / F'. \tag{7}$$

## *2.3 Complex amplitude in image plane*

The total distribution of complex amplitude $A_T(M', M'_O)$ created in the image plane by the multi aperture optical system can now be expressed as the sum of N amplitudes diffracted by the individual sub-pupils, centred on the paraxial image $M'_O$ and carrying phase terms $\phi_n$ proportional to the OPD defined in Eq. (4), i.e. $\phi_n = k\ \zeta_n$ where $k = 2\pi/\lambda$ and $\lambda$ is the wavelength of the electric field assumed to be monochromatic. Since all output pupils have the same diameter D', a general expression for $A_T(M', M'_O)$ is derived:

$$A_T(M',M'_O) = \hat{B}_{D'}(M'-M'_O)\, A(M',M'_O)\,, \tag{8}$$

where $\hat{B}_{D'}(M')$ stands for the amplitude diffracted by a single sub-pupil, and $A(M', M'_O)$ is a combination of complex amplitudes associated to the geometrical arrangement of the sub-pupils (each of them being reduced to a pinhole), that may be seen as a fictitious bi-dimensional diffraction grating:

$$A(M',M'_O) = \sum_{n=1}^{N} a_n \exp[i(\varphi_n + k\zeta_n)] = \sum_{n=1}^{N} a_n \exp[i\varphi_n]\exp\left[ik\left(\boldsymbol{M'P'_n} - m\,\boldsymbol{M'_O P_n}\right)/F'\right]. \tag{9}$$

Here $a_n$ and $\varphi_n$ respectively are the amplitude transmission factors and phase differences introduced along the $n^{th}$ arm of the interferometer. The phase terms $\varphi_n$ are suitable for introducing different optical delays on each individual arm, which is one of the basic principles of nulling interferometry, and the amplitudes $a_n$ are normalized such that $\sum_{n=1}^{N} a_n = 1$.

## 2.4 Object-Image relationship

Let us now consider a spatially extended sky object whose brightness distribution is described on-sky by the bi-dimensional distribution $O(\mathbf{s_O})$ that is rescaled into $O(M'_O)$ in the image plane. In the most general case, the multi-aperture system forms in the X'Y' plane an image I(M') composed of elementary intensity contributions $|A_T(M',M'_O)|^2$, weighted by function $O(M'_O)$:

$$I(M') = \iint_{M'_O} O(M'_O)\, |A_T(M',M'_O)|^2\, dM'_O$$

$$= \iint_{M'_O} O(M'_O) \left| \hat{B}_{D'}(M'\text{-}M'_O) \right|^2$$

$$\times \left| \sum_{n=1}^{N} a_n \exp[i\varphi_n] \exp\left[ik\left(\boldsymbol{M'P'_n} - m\,\boldsymbol{M'_O P_n}\right)/F'\right] \right|^2 dM'_O . \qquad (10)$$

At first glance, the previous expression looks somewhat different from the convolution relationship classically established between an object and its image formed by an optical system: that point will be further addressed in section 4.1.

## *2.5 Coupling into Single-Mode Waveguides*

Since the success of the FLUOR instrument [12] equipped with Single-Mode Fibers (SMF), the employment of SMWs in stellar interferometry has been growing extensively and is now being considered for future applications such as high-angular resolution imaging [13] or nulling interferometry [14]. SMW show the unique property of filtering the Wave-front Errors (WFE) introduced by the collecting optics (or the atmosphere for ground instrumentation) on larger spectral bands than simple pinholes do [15-16]. In addition they preserve the complex amplitude $A_O(M'_O)$ of the considered object. The relation between $A_O(M'_O)$ and the object irradiance distribution $O(M'_O)$ considered in the previous section is such that:

$$|A_O(M'_O)|^2 = O(M'_O). \qquad (11)$$

Hence the complex amplitude in the image field writes in the most general case:

$$A_I(M') = \iint_{M'_O} A_O(M'_O)\, A_T(M', M'_O)\, dM'_O , \qquad (12)$$

and the optical power coupled into a waveguide centered on point $M'_G$ in the X'Y' plane and whose modal function is noted G(M'), is equal to the square modulus of the so-called overlap integral [17]:

$$\rho(M'_G) = \iint_{M'} A_I(M')\, G^*(M'\text{-}M'_G)\, dM' \Big/ \gamma \tag{13}$$

with $\gamma = \left[ \iint_{M'} |G(M')|^2 \, dM' \right]^{1/2}$ and $^*$ denotes the complex conjugate. This study is limited to the case when the SMW is located on-axis (**O'M'$_G$ = 0**) and the sky object is an off-axis extra-solar planet centred on point $M'_P$ and described by the Dirac distribution $\delta(M'\text{-}M'_P)$. Then, combining Eqs. (12), (13) and (8) leads to:

$$\rho(M'_P) = \iint_{M'} A_T(M', M'_P)\, G^*(M')\, dM' \Big/ \gamma$$

$$= \iint_{M'} G^*(M')\, \hat{B}_{D'}(M'\text{-}M'_P)$$

$$\times \sum_{n=1}^{N} a_n \exp[i\varphi_n] \exp\left[ik\left(\boldsymbol{M'P'_n} - m\, \boldsymbol{M'_P P_n}\right)/F\right] dM' \Big/ \gamma \tag{14}$$

Further developments of the here above analytical formula are provided in section 5.

## 3 Roadmap to theory and numerical simulations

This brief section intended to help the reader is providing a short overview of the theory and numerical simulations that are presented afterwards. Two major, different cases will be considered successively:

- The Object-Image relationships applicable to various types of multi-aperture, high angular resolution imaging systems are first introduced and discussed in section 4. Different geometrical configurations of the entrance and exit sub-pupils characterized by their vectors $\boldsymbol{P_n}$ and $\boldsymbol{P'_n}$ have been studied. In particular, it is shown that for two typical sub-cases, namely the Super-Resolving Telescope (SRT) and the Axially Combined Interferometer (ACI), a third function F(M') appears in the classical Object-Image relationship, hereafter called the "far-field fringe function". For those two systems, the role of this function seems to be at least as important as the usual notions of Point Spread Function (PSF) and Optical Transfer Function (OTF). The imaging capacities of the SRT and ACI are further explored and illustrated by a set of numerical simulations.

- The same approach is also applied to nulling interferometers in section 5, where it serves for evaluating the throughput maps of the optical power emitted by an extra-solar planet, and coupled into a SMW. Here again, new simple analytical relationships (this time involving a far-field amplitude function and cross-correlation products) are derived. Numerical simulations allow to select the best combining scheme (axial or multi-axial), a major trade-off that is still open in the field of nulling interferometry.

Table 1 presents a synthetic view of the major studied cases. Most of them are further illustrated with the help of numerical simulations, whose main parameters such as input/output pupils geometry and optical characteristics are given in Table 2. All computations are carried out at a wavelength $\lambda$ = 10 µm. We consider "generic" collecting telescopes of 5-m diameter open at F/10. For all imaging configurations the focal lengths $F_C$ of the collimating optics were adjusted according to relation (1) in order to achieve a maximal densification in the exit pupil plane with a fast aperture number equal to 1. It must be noted that we imposed the same entrance baseline values B = 20 m for all Fizeau-like and axially combined nulling interferometric configurations. All those figures have not been optimized in depth, although they already provide good preliminary ideas of what real opto-mechanical implementations would be.

Table 1: Overall view of the considered optical configurations.

| Case | Number of telescopes | Sub-pupils configurations | Object-Image relationship | Section |
|---|---|---|---|---|
| Imaging configurations | | | | |
| Fizeau-like interferometer | | $\boldsymbol{P'_n} = m\, \boldsymbol{P_n}$ (golden rule $m' = m$, no densification) | Convolution | § 4.1 |
| Hypertelescope | 8 | $\boldsymbol{P'_n} = m'\, \boldsymbol{P_n}$ (high densification, $m/m' >> 1$) | No simplified expression | § 4.2 |
| Super-resolving telescope | 1 | $\boldsymbol{P_n} = \boldsymbol{0}$ | Convolution followed by multiplication with far-field fringe function | § 4.3 |
| Axially combined interferometer | 8 | $\boldsymbol{P'_n} = \boldsymbol{0}$ | Multiplication with far-field fringe function followed by convolution | § 4.4 |
| Nulling configurations | | | | |
| Nulling Fizeau-like interferometer | 2 (Bracewell) | $\boldsymbol{P'_n} = m\, \boldsymbol{P_n}$ | Not applicable | § 5.1 |
| Nulling super-resolving telescope | | $\boldsymbol{P_n} = \boldsymbol{0}$ | Not applicable | § 5.2 |
| Nulling axially combined interferometer | 2, 4 (Angel cross) and 8 | $\boldsymbol{P'_n} = \boldsymbol{0}$ | Not applicable | § 5.3 |

Table 2: Numerical values of main physical parameters for various simulation cases.

| Case | Number of entrance pupils | Number of exit pupils | B (m) | D (m) | F (m) | $F_C$ (mm) | B' (mm) | D' (mm) | F' (mm) | Section |
|---|---|---|---|---|---|---|---|---|---|---|
| Imaging configurations | | | | | | | | | | |
| Hypertelescope | 8 | 8 | variable | 5 | 50 | 300 | 60 | 30 | 100 | § 4.2 |
| Super-resolving telescope | 1 | 8 | 0 | 5 | 50 | 300 | 60 | 30 | 100 | § 4.3 |
| Axially combined interferometer | 8 | 1 | variable | 5 | 50 | 1000 | 0 | 100 | 100 | § 4.4 |
| Nulling configurations | | | | | | | | | | |
| Nulling Fizeau-like interferometer | 2 | 2 | 20 | 5 | 50 | 100 | 50 | 10 | 100 | § 5.1 |
| Nulling super-resolving telescope | 1 | 2 | 0 | 5 | 50 | 500 | 50 | 50 | 100 | § 5.2 |
| Nulling axially combined interferometer | 2, 4 and 8 | 1 | 20 | 5 | 50 | 100 | 0 | 10 | 100 | § 5.3 |

# 4 General imaging properties

In this section is demonstrated the basic property of multi-aperture imaging systems obeying to the golden rule (§ 4.1), followed by different theoretical expressions and numerical simulations undertaken for the cases of hypertelescopes (§ 4.2), super-resolving telescopes (§ 4.3) and axially combined, sparse apertures interferometer (§ 4.4).

## *4.1 Golden rule for Fizeau-like interferometers*

The famous "Pupil in = Pupil out" condition was initially introduced by Beckers *et al* [18-19], who were seeking to achieve an extended operational FoV on the Multiple Mirror Telescope (MMT) facility [20]. For that purpose they established that the "internal" and "external"

OPDs – herein the first and second terms of Eq. (7) – should cancel each other on the whole FoV, a condition that can only be realized if the exit pupil is homothetic to the entrance pupil. Alternative demonstrations of this statement, sometimes called "golden rule of stellar interferometry", can also be found in other papers [21-25]. Hereafter we will designate an interferometric array satisfying this golden rule as "Fizeau-like interferometer" (and not "Fizeau interferometer" since the latter appellation is sometimes understood as a monolithic telescope equipped with a multiple apertures screen).

The previous golden rule can be retrieved in a straightforward manner from the formalism used in Eq. (10). The condition for the input and output pupils to be homothetic just writes:

$$\boldsymbol{P'_n} = m' \boldsymbol{P_n} \tag{15}$$

for all individual sub-apertures ($1 \leq n \leq N$), $m'$ being the *geometrical* magnification factor of the entire multi-apertures array, from the input to the output pupil planes ($m'$ is also equal to B'/B, using the baseline parameters defined in Figure 2 and Figure 3). It is assumed in the whole study that $m'$ is a free parameter (not necessarily being equal to the optical magnification ratio $m$), which allows to study the cases of spaceborne, free-flying interferometers or hypertelescopes. Then the OPD $\zeta_n$ expressed in Eq. (7) becomes:

$$\zeta_n = (m' \boldsymbol{M'} - m \boldsymbol{M'_O}) \, \boldsymbol{P_n} \, / \mathrm{F'}, \tag{16}$$

and the "Pupil in = Pupil out" condition takes the simple form:

$$m' = m, \tag{17}$$

allowing the intensity distribution I(M') to become a convolution product between the object O(M') and the Point Spread Function of the multi-aperture optical system, itself being equal to the PSF of one individual sub-aperture multiplied by the far-field fringe function generated by the geometrical arrangement of the sub-pupils. Hence the Object-Image relationship of the Fizeau interferometer is finally applicable:

$$\mathrm{I(M')} = \iint_{\mathrm{M'_O}} \mathrm{O(M'_O)} \left| \hat{\mathrm{B}}_{\mathrm{D'}}(\mathrm{M'}\text{-}\mathrm{M'_O}) \right|^2 \left| \sum_{\mathrm{n}=1}^{\mathrm{N}} \mathrm{a_n} \exp[i\varphi_{\mathrm{n}}] \exp[i\mathrm{k}m(\boldsymbol{M'}-\boldsymbol{M'_O})\boldsymbol{P_n}/\mathrm{F'}] \right|^2 \mathrm{dM'_O}$$

$$= \mathrm{O(M')} * \left[ \left| \hat{\mathrm{B}}_{\mathrm{D'}}(\mathrm{M'}) \right|^2 \left| \sum_{\mathrm{n}=1}^{\mathrm{N}} \mathrm{a_n} \exp[i\varphi_{\mathrm{n}}] \exp[i\mathrm{k}m\boldsymbol{M'}\boldsymbol{P_n}/\mathrm{F'}] \right|^2 \right], \qquad (18)$$

which is in agreement with Harvey *et al* [26]. Our current knowledge of Fizeau-like interferometers is today well established: their essential property (dictated by the golden rule) is that their full output pupil (in plane $X'_P Y'_P$) must be a reduced replica of their entrance pupil (in plane $X_P Y_P$) as shown in Figure 4. Harvey *et al* [27-28] demonstrated that in that case the best images are obtained when the “dilution factor” is maximized, i.e. when two or more input/output sub-pupils are placed edge to edge, providing a better OTF plane coverage. Until now, most of the multi-apertures imaging systems that have been constructed are Fizeau-like interferometers (e.g. the Multiple Mirror Telescope [20], the Large Binocular Telescope [29] or the Multiple Instrument Distributed Aperture Sensor [30]).

We shall now focus our attention on three attractive cases where the golden rule is not respected – as in the original Michelson apparatus. The studied geometrical configurations for the entrance and exit apertures are depicted in Figure 2, while the values of the major parameters used for the numerical simulations are those provided in Table 2.

## 4.2 Hypertelescopes

The major difference between the previous Fizeau-like interferometer and the hypertelescope concept originally proposed by Labeyrie in ref. [7] is that the golden rule is no longer respected. Hence the convolution relation (18) is not applicable and the classical notions of PSF and OTF acquire a different signification. It has been shown, however, that a hypertelescope is still able to provide direct, highly spatially resolved images of extra-solar planets in a narrow FoV when kilometric baselines B are imposed [7]. The conceptual optical layout of the system is summarized on the Figure 5 that shows, when compared with the basic design of Figure 3, an additional group of lens (or mirrors) incorporated along each separated arm. That group is named “beam densifier” and is composed of three optical components L4_n, L5_n and L6_n. The couple (L4_n; L6_n) has the principal function of enlarging the diameter D’ of the N output pupils, in such a way that they are re-arranged side by side (or as close as possible) in the plane of the recombining optics (see the Figure 4): it has been demonstrated that such “pupil densification” allows to minimize the core of the PSFs with respect to the classical Fizeau configuration, and therefore to improve the spatial resolution of the images [11]. Optional diverging optics L5_n can serve to relay the pupil images downstream. It can be assumed without loss of generality that the input and output focal lengths $F_R$ of the whole relay optics (from L3_n to L4_n) are identical, and thus their magnification is taken equal to 1.

From a theoretical point of view, the hypertelescope is often characterized by its “densification factor” *d* (obviously linked to Harvey’s dilution factor evoked the previous section), here equal to *m* / *m*’ according to the employed notations. The OPD $\zeta_n$ can thus be rewritten as:

$$\zeta_n = m' (\boldsymbol{M'} - d\, \boldsymbol{M'_O})\, \boldsymbol{P_n} / \mathrm{F'} \tag{19}$$

and it can be expected that for very long baselines (e.g. B > 1 km), *d* becomes significantly higher than unity, therefore the vector ***M’*** can be neglected in Eq. (19). Hence the hypertelescope would tend to behave like the axially combined interferometer described in section 4.4. That point, however, has not been confirmed by the results of the numerical simulations presented here, for which is considered a free-flying array composed of eight telescopes with varying baseline B, and whose apertures are disposed along a square contour as shown on the left top panel of Figure 2. Owing to the values adopted here for F and $F_C$ (see Table 2), the golden rule is respected when $m = m' = B' / B = F_C / F = 3 / 500$. This condition leads to an entrance baseline B equal to 10 m when the baseline of the exit pupils B' is set equal to 60 mm in order to achieved a maximal densification (see Table 2). The point is illustrated by Figure 6, showing different simulated images of a given object (here a picture of Saturn, not to scale) formed by a hypertelescope for various values of B. It is observed that the best image resolution is clearly achieved when the golden rule is fulfilled (i.e. B = 10 m implying that the eight entrance pupils are connected), while for longer baselines the images get perturbed by destructive interference patterns without showing appreciable resolution enhancement. Mathematically, this is most probably due to the fact that the high densification factors associated with very long baselines B are actually used to probe small angular size objects: hence both vectors $\boldsymbol{M'}$ and $d\,\boldsymbol{M'_O}$ remain of the same magnitude order and none them can be neglected, preventing Eq.(10) from being reducible to a convolution product. It can be concluded that the classical golden rule remains fully applicable to hypertelescopes, which should noticeably restrict their scientific domain of application: in fact, a hypertelescope governed by the golden rule is nothing else than a Fizeau-like interferometer such as described in the previous section, and will suffer from the same limitations for highly diluted arrays (e.g. spurious parasitic images superimposed to the observed sky-object [11] [28]).

It has also to be highlighted that the here above numerical computations were long and cumbersome (15 hours of computing time required for a 149 × 149 image sampling), since the integral in Eq. (10) was evaluated iteratively for each grid sample. This drawback disappears when Eq. (10) can be reduced to a convolution product, which happens for two particular cases (in addition to the Fizeau-like interferometer) that are examined in the following sections: the super-resolving telescope (§ 4.3) and the axially recombined interferometer (§ 4.4).

## *4.3 Super-Resolving Telescope (SRT)*

The term super-resolving telescope is inspired from Toraldo di Francia [31], who showed that single pupil optical systems may attain sub-diffraction resolution when their surface is constituted of alternating concentric rings of variable thickness and phase differences. The principle has been demonstrated experimentally in the microwave band [32], initiating discussions to assert if the Rayleigh limit was overcome – indeed it was not, since in that case most of the optical power is radiated outside of the first lobe of the Airy spot. More recently, Greenaway and Spaan evoked the "pupil replication" technique [33-34], which is not very far from the principle presented here below. Nevertheless, their formalism was limited to the one-dimensional case, which probably prevented them from deriving the general Object-Image relationship (20a) applicable to the SRT.

Mathematically speaking, a super-resolving telescope is obtained when all the individual entrance sub-apertures of a hypertelescope are merged into the single pupil of one monolithic telescope (i.e. $\boldsymbol{P_n} = \mathbf{0}$ whatever is n). Practically, this can be realized by the schematic optical layout presented in Figure 7: one single collecting afocal telescope optically feeds a number N of off-axis, parallel exit arms that are multi-axially recombined downstream by the fast aperture lens or mirror L'. The beams are separated by means of a set of cascaded beamsplitters noted BS1 and BS2_n placed at the output port of the afocal telescope. The whole optics arrangement is such that each beam experiences the same number of reflections and transmissions on the beamsplitters and folding mirrors M1_n (this requirement may not be necessary for direct imaging, but will become crucial in the perspective of a nulling SRT such as proposed in § 5.2). The unused reflected or transmitted beams are directed towards metrology sensors that can be used for example to monitor the telescope pointing misalignments or wave-front errors. The beams densifiers and combining optics are similar to those already described in the previous section, and the pupil conjugations are ensured by either L2 or L5_n diverging elements, or both. It must be emphasized that all the optical components comprised between BS1 and the exit recombiner may be of rather modest size and assembled into a common structure, thus relaxing considerably the mechanical and thermal stability requirements applicable to the free-flying hypertelescope.

Setting $\boldsymbol{P_n} = \mathbf{0}$ in Eq. (10) readily conducts to a simplified expression of the image distribution I(M') in the X'Y' plane, being equal to the convolution product between the object and the PSF of a sub-pupil, multiplied by a masking function F(M') – herein called far-field fringe function – resulting from constructive and destructive interferences generated by the geometrical disposition of the exit sub-apertures:

$$\mathrm{I(M')} = \iint_{\mathrm{M'_O}} \mathrm{O(M'_O)} \left|\hat{\mathrm{B}}_{\mathrm{D'}}(\mathrm{M'-M'_O})\right|^2 \left|\sum_{n=1}^{N} \mathrm{a_n} \exp[i\varphi_\mathrm{n}] \exp[i\mathrm{k}\,\boldsymbol{M'}\boldsymbol{P'_n}/\mathrm{F'}]\right|^2 \mathrm{dM'_O}$$

$$= \mathrm{F(M')}\left[\mathrm{O(M')} * \left|\hat{\mathrm{B}}_{\mathrm{D'}}(\mathrm{M'})\right|^2\right], \tag{20a}$$

where

$$\mathrm{F(M')} = \left|\sum_{n=1}^{N} \mathrm{a_n} \exp[i\varphi_\mathrm{n}] \exp[i\mathrm{k}\,\boldsymbol{M'}\boldsymbol{P'_n}/\mathrm{F'}]\right|^2 \tag{20b}$$

The previous relationships evidence the important role of the far-field fringe function. Figure 8 displays a grey-scale transmission map and a horizontal slice of F(M'), which can be considered as an occulting screen pierced by a square grid of transparent holes and masking the observed object. The mask is indeed the Fourier transform of the sum of eight individual Dirac distributions whose locations correspond to the centres of each output sub-pupil (see

Figure 2), and appears as a series of thin transmission peaks arranged on a regular square grid. Examining the slice of F(M') at the bottom of Figure 8, we find that the peak width is significantly narrower than the width of the PSF of an equivalent 5-m telescope for the considered wavelength (the two functions are respectively represented by thick and thin lines at the bottom of Figure 8). One can then imagine to introduce small misalignments of the telescope optical axis with respect to the sky object, allowing to scan it spatially as if it was observed through a moving mask. The accumulated images could then be recombined via a shift-and-add mosaïcking procedure whose principle is described in Figure 9: starting from an arbitrary transmission peak, the square angular area separating it from its closest neighbors (140 milli-arcsecs wide, as indicated by dotted lines in Figure 9) is explored along the U and V axes by steps of 30 milli-arcsecs, which is roughly equal to the full width at half maximum of the individual transmission peak. All the acquired images are then stored, recentred and added incoherently, yielding an apparently super-resolved image where the information present in the differently masked objects was simply combined. However, and regardless of the much longer required telescope observation time, the method suffers from two fundamental limitations:

1) The essential limitation is indeed a natural consequence of basic relation (20a), where the convolution product between the object O(M') and the PSF of the telescope $|\hat{B}_{D'}(M')|^2$ takes place *before* the super-resolving process is started: it may thus be expected that a large amount of the spatial information regarding the sky-object has already disappeared, and will not be retrieved by means of the sole far-field fringe function F(M') and its associated mosaïcking procedure.
2) Moreover, for this particular, eight sub-apertures configuration, the function F(M') exhibits regular parasitic peaks of 25 % transmission (clearly visible in Figure 8), thereby introducing spatial crosstalk between the successive elementary acquisitions and a scrambling of the final, reconstructed image. However this drawback seems to be less critical than the previous one, since Fourier optics theorems tell us that incorporating more output sub-pupils to the SRT should minimize and even eliminate the parasitic peaks.

From a purely computational point of view, the convolution product in Eq. (20a) can be quickly and efficiently calculated by means of conventional, fast double Fourier transform algorithms. In Figure 10 are presented a series of numerical simulations illustrating the whole measurement process: top row shows the same sky object as in Figure 6 and its image observed through a 5-m telescope at the wavelength $\lambda$ = 10 $\mu$m. The bottom left panel exhibits a raw image produced by a SRT of same diameter having eight exit sub-apertures and whose geometrical characteristics are provided in Table 2: it appears as a series of thin dots disposed on a regular grid pattern, whose intensities are proportional to the brightness of the extended celestial object (note the presence of the faint parasitic transmission peaks mentioned here above). The bottom right panel of Figure 10 depicts the result of a crude $4 \times 4$ mosaïcking algorithm, showing no real significant improvement in angular resolution of the observed object with respect to the image formed by the traditional monolithic telescope, although the general appearance of the image has been significantly altered. Hence it can be concluded that, even if it obeys to an unconventional Object-Image relationship, the here above presented system does not show plain super-resolution capacities. Its major advantage, however, is to concentrate the luminous energy emitted from celestial objects onto very small sensing areas of the detection plane, corresponding the peaks of the far-field fringe function. This basic property will serve as the starting point for the concept of nulling super-resolving telescope presented in section 5.2.

## *4.4 Axially Combined Interferometer (ACI)*

The technique of axial (or coaxial) recombination for stellar interferometry has been known for a long time, even if a majority of existing facilities or instruments rather uses multi-axial combining. Nowadays axial recombination is considered as a major scheme for nulling interferometry, in the frame of which important efforts are being undertaken to design very symmetrical optical layouts [35-36]. An example of such an arrangement is shown in Figure 11: as in the case of the SRT, it involves an equal number of reflective and transmissive interfaces on the beamsplitters and fold mirrors along each interferometer arm, and allows the implementation of metrological sensors for OPD and tip-tilt measurements. The collecting optics are not shown in Figure 11, since they are strictly identical to those of the hypertelescope (section 4.2, Figure 5).

The mathematical expression of an image I(M') created at the focal plane of an axially combined interferometer is derived from Eq. (10), assuming that all output apertures are superimposed, i.e. $\boldsymbol{P'_n} = \mathbf{0}$ whatever is n. Here again I(M') reduces to a convolution product that can be computed accurately and rapidly:

$$\mathrm{I(M')} = \iint_{\mathrm{M'_O}} \mathrm{O(M'_O)} \left|\hat{\mathrm{B}}_{\mathrm{D'}}(\mathrm{M'\text{-}M'_O})\right|^2 \left|\sum_{n=1}^{N} \mathrm{a_n} \exp[i\varphi_n] \exp[-i\mathrm{k}\, m\boldsymbol{M'_O}\boldsymbol{P_n} / \mathrm{F'}]\right|^2 \mathrm{dM'_O}$$

$$= [\mathrm{O(M')\,F(M')}] * \left|\hat{\mathrm{B}}_{\mathrm{D'}}(\mathrm{M'})\right|^2, \qquad (21a)$$

with

$$\mathrm{F(M')} = \left|\sum_{n=1}^{N} \mathrm{a_n} \exp[i\varphi_n] \exp[-i\mathrm{k}\, m\boldsymbol{M'}\boldsymbol{P_n} / \mathrm{F'}]\right|^2 . \qquad (21b)$$

We find that I(**M'**) is now equal to a convolution product involving the PSF of the sub-pupil, on the one hand, and the multiplication product of the object with the far-field fringe pattern F(M’) generated by the entrance pupils arrangement, on the other hand. It is quite remarkable that, for the main cases considered in section 4, namely the Fizeau-like interferometer, SRT and ACI, the resulting image distribution involves the same three bi-dimensional functions (the object, the PSF of an individual sub-pupil and the far field fringe function), linked together by multiplication and convolution operators in subtle different order. The ACI imaging capacities are illustrated in Figure 12, showing another example of sky-object (left top panel), its image when observed through a single telescope (right top panel), and the way it would be revealed at the image plane of an ACI composed of eight collecting telescopes, for increasing values of the entrance baseline B. For the shortest baseline B = 10 m (left bottom panel) the image is scrambled by the function F(M’), but this effect gradually vanishes when the baseline is enlarged (right bottom panel). Moreover, it has been noticed that the general image aspect does not improve significantly beyond B = 20 m. When considering the here above Object-Image relationship (21a) applicable to the ACI concept, the previous results may be explained as follows:

- For short baselines B, the angular separation between two neighboring transmission peaks of the far-field fringe function (as displayed in Figure 8) is relatively large. F(M’) acts as a mask sampling the observed sky object with a degraded angular resolution.
- For longer baselines the resolution becomes limited by the PSF of an individual sub-pupil projected on-sky – the function $|\hat{B}_{D'}(\mathbf{M'})|^2$. No further improvement occurs when the angular separation of the transmission peaks in F(M’) is smaller than the PSF width.

It finally turns out that for very long baselines, the imaging properties of the ACI are similar to those provided by a single individual telescope: in other words, the global resolving power is ultimately limited by the diffraction lobe of the individual sub-apertures, and only a gain in radiometric performance may be expected from the ACI concept.

To conclude this already long section, it is recalled that the Fourier optics formalism presented herein makes it possible to express the image distributions formed by some typical high angular resolution systems under a simple form involving convolution products. Three important cases have been distinguished: the well known Fizeau-like interferometer, a candidate super-resolving telescope, and an axially combined interferometer inspired from the hypertelescope concept. It was demonstrated that the two last types of systems are not governed by the classical Object-Image relationship, but that they nevertheless do not seem to present extreme resolving capacities. The three major concepts will now be re-examined in the following section that deals with their application in the framework of nulling interferometry, making use of a very similar formalism.

## 5 Application to nulling interferometry

Nulling interferometry [4-6] is nowadays a widely known and studied technique: it aims at discovering Earth-like planets orbiting around nearby stars and characterizing their atmospheres in hope of recognizing signs of life. Because the searched planets are very close and much fainter than their parent star, the technical requirements are far more difficult to meet than in direct imagery – say, by two or three orders of magnitude. During the last decade, the European Space Agency (ESA) and National Aeronautics and Space Administration (NASA) extensively developed two major projects of nulling interferometers, respectively named Darwin [37] and TPF-I (Terrestrial Planet Finder Interferometer [38]). Hence the quest for extra-solar planet could finally become the major astronomical challenge of the 21st century.

Practically, any of the high angular resolution systems described in section 4 could be transformed into a nulling instrument, provided that one or several Achromatic Phase Shifter (APS) devices producing a $\pi$ phase shift between a couple of optical trains are added within the optical layout. The recent manufacturing and tests of high-performance infrared APS have recently been reported [39]. Nulling interferometry could also benefit from current progress on single-mode waveguides technology [14]: one very popular interferometer design, named "fibered nuller", consists indeed in illuminating the core of a single-mode fiber with two or more off-axis beams being in phase opposition – i.e. $\varphi_n = 0$ or $\pi$ in Eq. (14). It must be emphasized that two of the deepest nulling ratios ever obtained in the optical laboratory just exploited that technique [40-41]. An example of interferometer configuration incorporating a nulling periscope APS and a SMF centred on the origin O' of the X'Y' plane is therefore depicted in Figure 13 (collecting and densifying optics are not shown). The main scope of the following paragraphs is to derive simplified expressions of the so-called extinction or nulling maps of the interferometer characterizing the whole destructive and constructive fringe pattern projected on-sky [9-10]. Those maps are normalized such that their numerical values are directly equal to the actual instrument throughput as a function of the angular position of the planet: obviously the nulling map must always be equal to zero on-axis, since this is the theoretical direction of the parent star. The numerical results are then analyzed and compared together in order to define the most efficient recombination scheme (Fizeau-like, multi-axial or axial), a major trade-off that is still open in the framework of the Darwin and TPF-I projects.

## 5.1 *Nulling Fizeau-like interferometer*

Let us first consider an eight apertures stellar interferometer satisfying the golden rule of stellar interferometry, and transform it into a nulling interferometer. This is realized by means of a series of achromatic, $\pi$ phase-shifters arranged on the exit pupils as shown on the left bottom panel of Figure 2 (since only the case of the Bracewell-like configuration is considered there, see below). Inserting relations (15-17) into Eq. (14) and assuming that both functions G(M') and $\hat{B}_{D'}(M')$ are real and centro-symmetric, which is true as long as no optical aberrations or manufacturing errors are introduced within the system, then enables to express the overlap integral ρ(M') as the cross correlation product of Eq. (22): here the far-field fringe function defined in section 4 has been replaced by its equivalent in the SMW formalism, namely the "far-field amplitude function" that is the linear combination of the complex amplitudes generated by the sub-apertures arrangement:

$$\rho(M') = G(M') \otimes \left[ \hat{B}_{D'}(M') \sum_{n=1}^{N} a_n \exp[i\varphi_n] \exp[-ikm\boldsymbol{M'}\boldsymbol{P_n} / F'] \right] \Big/ \gamma , \tag{22}$$

where symbol ⊗ denotes the cross correlation product. By convention, the global throughput of the instrument T(M') including the coupling efficiency into the single-mode fiber is finally estimated as:

$$T(M') = |\rho(M')|^2 \times P / P_0, \tag{23}$$

P being the total power coupled into the SMF, and $P_0$ the total energy radiated from the planet and collected by the whole entrance pupil of the interferometer. Gray-scale representations of the obtained distribution T(M') are depicted in the top row of Figure 14, and the main achieved performance in terms of planet throughput and Inner Working Angle (IWA) is summarized in the Table 3: as used for coronagraphs, the IWA is defined as the minimal angular distance from the star at which the planet throughput exceeds 50 % of its maximal value in the whole FoV.

Table 3: Summary of nulling interferometers parameters and achievable performance (see also Table 2).

| Case | N | Planet throughput | Inner Working Angle (IWA) | SMF core radius |
|---|---|---|---|---|
| Nulling Fizeau-like interferometer | 2 | 0.5 % | 83 milli-arcseconds | 10.4 μm |
| Nulling SRT | 2 | 4.8 % | 154 milli-arcseconds | 8.4 μm |
| Nulling ACI (Bracewell) | 2 | 77.9 % | 26 milli-arcseconds | 84.1 μm |
| Nulling ACI (Angel cross) | 4 | 75.0 % | 36 milli-arcseconds | 84.1 μm |
| Nulling ACI (8 telescopes) | 8 | 64.7 % | 63 milli-arcseconds | 84.1 μm |

The numerical computation shows that the maximal achieved throughput for the planet is only 0.5 %, which seems very low and could in practice only be counterbalanced by prohibitive observation times. Moreover, the results achieved for the four-telescopes (Angel cross) and eight telescopes configurations defined in Figure 5 are so dramatically worse (i.e. significantly inferior to 0.1 %) that they are even not given in the Table. The point may be interpreted as follows: since the golden rule is respected, it ensures a certain uniformity of the OPD $\zeta_n$ within the entire Field of View. Here however, the phase shifts $\varphi_n$ have been adjusted so that a nulled, destructive fringe is created at the FoV centre. One could therefore argue that the destructive interference spreads through the whole FoV with the unwanted consequence of minimizing the planet throughput everywhere. Hence the golden rule for stellar interferometry would indeed be detrimental to nulling interferometers. Nevertheless, a rigorous demonstration of the latter statement is not straightforward: we may therefore consider it as a rule of thumb deserving future studies and explanations.

## *5.2 Nulling super-resolving telescope*

The principle of the nulling SRT has been proposed in a recent communication [42]: it is indeed a super-resolving telescope similar to those discussed in section 4.3, where a number of APS are added into each optical arm before the exit recombining optics (Figure 7). The mathematical expression of the overlap integral ρ(M') is derived from Eq. (14) assuming that $\boldsymbol{P_n} = \mathbf{0}$ whatever is n, corresponding to the case when all entrance apertures are merged. Then the expression of ρ(M') becomes:

$$\rho(\mathrm{M'}) = \left[ \mathrm{G(M')} \sum_{n=1}^{N} \mathrm{a_n} \exp[i\varphi_n] \exp[-i\mathrm{k}\boldsymbol{M'P'_n} / \mathrm{F'}] \right] \otimes \hat{\mathrm{B}}_{\mathrm{D'}}(\mathrm{M'}) \Big/ \gamma \qquad (24)$$

which is very similar to relation (22), the functions G(M') and $\hat{\mathrm{B}}_{\mathrm{D'}}(\mathrm{M'})$ having just been swapped. An example of throughput map T(M') is displayed on the second row of Figure 14. Here the maximal throughput of the planet for the basic configuration including two exit symmetric arms is found equal to 4.8 %, which is 10 times higher than for the nulling Fizeau-like interferometer, but still remains insufficient (it has been checked here also that the throughput is not acceptable for configurations involving a higher number of optical arms). Here the point seems to be related to the maximal size of the fiber core (that has to be lower than 8.4 μm in order to transmit the sole fundamental mode at 10 μm), on the one hand, and to the angular area of the central null increasing as extra exit arms are added, on the other hand: hence a centred, on-axis SMF will collect less and less photons as the nulling area is extended. One solution to improve the throughput could be to decenter the SMF, or eventually to implement a SMW array if that technology becomes available.

## *5.3 Nulling axially combined interferometer*

We finally examine the case of a nulling, axially combined interferometer that can simply been extrapolated from the ACI design described in section 4.4 with the addition of N APS along all the interferometer arms of Figure 11. When constituted of only two collecting telescopes, this nulling ACI is nothing else than the Bracewell's original design [4]. Imposing that $\boldsymbol{P'_n} = \mathbf{0}$ whatever is n readily leads to the following expression of ρ(M'):

$$\rho(M') = \left[ \sum_{n=1}^{N} a_n \exp[i\varphi_n] \exp[ik m \boldsymbol{M'} \boldsymbol{P_n} / F'] \right] \left[ G(M') \otimes \hat{B}_{D'}(M') \right] \Big/ \gamma . \qquad (25)$$

With respect to Eq. (24), $\hat{B}_{D'}(M')$ has been permuted with the far-field amplitude function (here it must be noticed that the previous relationship is in accordance with those already published in Refs. [9-10] that were precisely limited to the case of axial recombination). Numerical simulations based on Eq. (25) were then carried out for different interferometric arrays, respectively based on two, four and eight telescopes (see Figure 2), and led to the very satisfactory results reported in Table 3, showing that the estimated throughput always exceeds 60 %, even for the eight telescopes configuration having the most extended nulling area (see Figure 14). Hence the present section devoted to fibered nulling interferometers finally evidences a marked superiority of the axial combining scheme with respect to other designs (i.e. nulling Fizeau interferometer or SRT). Therefore it seems that the question of axial or multi-axial recombining optics, which remains one of the major open tradeoffs in nulling interferometry, may be given here an element of answer.

## 6 Future work

The perspective of detecting terrestrial extra-solar planets by means of a nulling fibered ACI deserves additional discussions and future work devoted to some specific issues that are briefly summarized below. Other studies may also be undertaken in order to better assess the imaging capacities of the SRT and the ACI.

**1) Chromatic dispersion**

Whether intended for nulling or imaging purposes, one common feature to interferometric arrays is their strong dependence on chromatic dispersion, since the angular scale of both functions $|\hat{B}_{D'}(\mathbf{M'})|^2$ and F(M') defined in section 4 is directly proportional to the wavelength $\lambda$. This difficulty might be overcome by inserting Wynne compensators into the combining optics [43]: this type of system is especially designed to present a lateral chromatism that is inversely proportional to $\lambda$, hence the final diffraction pattern in the image plane should be free of chromatism. However this solution requires to integrate dioptric components within the optical layout, which may induce severe practical constraints in the thermal infrared spectral band selected for the space missions Darwin and TPF-I. An alternative scheme inspired from modern integral field spectroscopy may consist in a reflective image transformer [44] placed downstream a diffraction grating, further rescaling the individual spectral images and superimposing them in the final focal plane. It must be highlighted that the practical realization of such "inverted image slicers" only requires mature technologies, as confirmed by recent publications [45-46].

**2) Radiometric performance**

Some preliminary estimations of the radiometric efficiency of the three presented nulling interferometers have already been provided in section 5 (see Table 3). This work would naturally need to be completed for the other mentioned systems (i.e. imaging hypertelescope, SRT and ACI). More generally, a complete Signal-to-Noise Ratio (SNR) budget should be established for each considered case. Those SNR budgets should take into account various parameters such as the magnitude of the observed celestial object, total light collecting area, integration time (that mat attain several days for space instruments) and different types of noise characterizing modern detector systems (e.g. photon noise, read-out noise and dark current).

**3) Pupil imaging requirements**

It has been highlighted in section 2.2 that the assumption of entrance sub-pupils being re-imaged on their associated exit sub-pupil is fundamental, since it allows to derive the relationship (7) on which the remainder of the theory is based. In practice this condition dictates the implementation of pupil relaying optics and possibly of delay lines in order to equalize the OPDs in all different sub-apertures. Here the quality of the pupil imaging and the pistons, shear and defocus errors leading to imperfect sub-pupils matching should be of prime importance for the quality of the achieved nulling or imaging performance. Therefore a huge effort in optical design and tolerancing analysis in view of defining quantitative requirements remains to be carried out. Here it must be noticed that the condensed form of Eq.(7) makes it suitable for introducing some of the mentioned defects.

**4) Entrance and exit pupil configurations**

In the whole paper were only considered square, redundant input and output optical arrays that were found satisfactory for most nulling or imaging cases. It is well-known however that such configurations are not optimal for the Fizeau-like interferometers, and that some other ones (e.g. Golay or circular, non-redundant arrays) provide better coverage of the OTF plane and a consequent image enhancement. It would be of prime interest to verify if this conclusion remains valid for the herein presented SRT and ACI obeying to different Object-Image relationships.

**5) Optical system modeling**

The frame of this study was from the beginning restricted to first-order optics and scalar diffraction theory, which seem reasonable hypotheses when dealing with low angular aperture optical systems. But it must be noticed that most of the herein presented systems (i.e. hypertelescope and nulling or imaging SRT) make use of fast aperture recombining optics, thus at least in their cases a vectorial diffraction analysis seems mandatory.

## 7 Summary

In this paper were reviewed some general properties of various advanced, multi-aperture optical systems dedicated to direct, high angular resolution imaging or to the detection and characterization of extra-solar planets with the help of nulling interferometry technique. The use of a rather simple Fourier optics formalism applicable to both imaging arrays and nulling interferometers enabled to express those imaging and nulling capacities as convolution or cross correlation products suitable for fast and accurate numerical computing. A variety of high angular resolution systems were considered, and in my view the preliminary conclusions of this theoretical study are twofold:

- The axial combination scheme seems to be the most recommendable for a multi-aperture, fibered nulling interferomer, at least from the point of view of radiometric efficiency. This conclusion may have some important consequence on the architecture and design of the whole free-flying telescope array.
- Two of the presented optical systems, namely the imaging super-resolving telescope and axial combining interferometer, are governed by non classical Object-Image relationships that may be appended to Fourier optics theory. However their angular resolution seems to be ultimately limited by the diffraction lobes of an individual collecting telescope.

To conclude, it is recalled that the entire study presented in this paper is only valid in the frame of first-order geometrical optics and scalar diffraction theory, applied to monochromatic light waves. Furthermore, it is assumed that all entrance sub-apertures are optically conjugated with their associated exit pupils, and that no pupil aberrations exist (although piston and pupil decentring errors could be easily introduced in the present formalism). Also, no “real world” constraints such as manufacturing, aligning and testing feasibility, instrumental biases, detector noises or atmospheric seeing were covered in this purely theoretical work. It is likely, however, that the herein described high angular resolution systems should preferably be envisaged for space applications.

The author would like to thank his colleagues D. Mourard and Y. Rabbia for inspiring discussions about the golden rule, hypertelescopes and nulling interferometry.

## REFERENCES

1. A. A. Michelson and F. G. Pease, "Measurement of the diameter of alpha Orionis with the interferometer," Astrophys. J. vol. 53, p. 249-259 (1921).
2. A. Labeyrie, "Interference fringes obtained on Vega with two optical telescopes," Astrophys. J. vol. 196, p. L71-L75 (1975).
3. P. R. Lawson, Selected papers on long baseline stellar interferometry, SPIE Milestones Series vol. MS 139 (1997).
4. R. N Bracewell and R. H. MacPhie, "Searching for non solar planets," Icarus vol. 38, p. 136-147 (1979).
5. J. R. Angel, "Use of a 16 m telescope to detect earthlike planets," Proceedings of the Workshop on the Next Generation Space Telescope, P. Bely and C.J. Burrows eds. (Space Telescope Science Institute, Baltimore, Md., 1990), p. 81–94.
6. J. R. Angel, J. H. Burge and N. J. Woolf, "Detection and spectroscopy of exo-planets like Earth," Proceedings of the SPIE vol. 2871, p. 516-519 (1997).
7. A. Labeyrie, "Resolved imaging of extra-solar planets with future 10-100 km optical interferometric arrays," Astronomy and Astrophysics Supplement Series vol. 118, p. 517-524 (1996).
8. O. P. Lay, "Imaging properties of rotating nulling interferometers," Applied Optics vol. 44, p-5859-5871 (2005).
9. F. Hénault, "Computing extinction maps of star nulling interferometers," Optics Express vol. 16, p. 4537-4546 (2008).
10. F. Hénault, "Fine art of computing nulling interferometer maps," Proceedings of the SPIE vol. 7013, n° 70131X (2008).

11. O. Lardière, F. Martinache and F. Patru, “Direct imaging with highly diluted apertures - I. Field of view limitations,” Monthly Notices of the Royal Astronomical Society vol. 375, p. 977-988 (2007).
12. V. Coudé du Foresto, S. Ridgway and J. M. Mariotti, “Deriving object visibilities from interferograms obtained with a fiber stellar interferometer,” Astron. Astrophys. Suppl. Ser. Vol. 121, p. 379–392 (1997).
13. G. Perrin, S. Lacour, J. Woillez and E. Thiébaut, “High dynamic range imaging by pupil single-mode filtering and remapping,” Monthly Notices of the Royal Astronomical Society vol. 373, p. 747-751 (2006).
14. A. Ksendzov, O. Lay, S. Martin, J. S. Sanghera, L. E. Busse, W. H. Kim, P. C. Pureza, V. Q. Nguyen, I. D. Aggarwal, “Characterization of mid-infrared single mode fibers as modal filters,” Applied Optics vol. 46, p. 7957-7962 (2007).
15. C. Ruilier and F. Cassaing, “Coupling of large telescope and single-mode waveguides,” J. Opt. Soc. Am. A vol. 18, p. 143-149 (2001).
16. M. Ollivier and J.M. Mariotti “Improvement in the rejection rate of a nulling interferometer by spatial filtering,” Applied Optics vol. 36, p. 5340-5346 (1997).
17. R. E. Wagner and W. J. Tomlinson, “Coupling efficiency of optics in single-mode fiber components,” Applied Optics vol. 21, p. 2671-2688 (1982).
18. J. M. Beckers, E. K. Hege and P.A. Strittmatter, “Optical interferometry with the MMT,” Proceedings of the SPIE vol. 444, p. 85-92 (1983).
19. J. M. Beckers, “Field of view considerations for telescope arrays,” Proceedings of the SPIE vol. 628, p. 255-260 (1986).
20. E. K. Hege, J. M. Beckers, P. A. Strittmatter, D. W. McCarthy, “Multiple mirror telescope as a phased array telescope,” Applied Optics vol. 24, p. 2565-2576 (1985).

21. A. B. Meinel, “Aperture synthesis using independent telescopes,” Applied Optics vol. 9, p. 2501-2504 (1970).

22. W. A. Traub, “Combining beams from separated telescopes,” Applied Optics vol. 25, p. 528-532 (1986).

23. F. Merkle, “Synthetic-aperture imaging with the European Very Large Telescope,” J. Opt. Soc. Am. A vol. 5, p. 904-913 (1988).

24. L. D. Weaver, J. S. Fender, C. R. de Hainaut, “Design considerations for multiple telescope imaging arrays,” Optical Engineering vol. 27, p. 730-735 (1988).

25. M. Tallon and I. Tallon-Bosc, “The object-image relationship in Michelson stellar interferometry,” Astron. Astrophys. vol. 253, p. 641-645 (1992).

26. J. E. Harvey, A. B. Wissinger, A. N. Bunner, “A parametric study of various synthetic aperture telescope configurations for coherent imaging applications,” Proceedings of the SPIE vol. 643, p. 194-207 (1985).

27. J. E. Harvey, C. Ftaclas, “Field-of-view limitations of phased telescope arrays,” Applied Optics vol. 34, p. 5787-5798 (1995).

28. J. E. Harvey, A. Kotha, R. L. Phillips, “Image characteristics in applications utilizing dilute subaperture arrays,” Applied Optics vol. 34, p. 2983-2992 (1995).

29. E. E. Sabatke, J. H. Burge, P. Hinz, “Optical design of interferometric telescopes with wide fields of view,” Applied Optics vol. 45, p. 8026-8035 (2006).

30. D. M. Stubbs, A. L. Duncan, J. T. Pitman, R. D. Sigler, R. L. Kendrick, J. F. Chilese, E. H. Smith, “Multiple instrument distributed aperture sensor (MIDAS) science payload concept,” Proceedings of the SPIE vol. 5487, p. 1444-1452 (2004).

31. G. Toraldo di Francia, “Super-gain antennas and optical resolving power,” Nuovo Cimento vol. 9, p. 426-438 (1952).

32. A. Ranfagni, D. Mugnai and R. Ruggeri, "Beyond the diffraction limit: Super-resolving pupils," Journal of Applied Physics vol. 95, p. 2217-2222, (2004).
33. A. H. Greenaway, F. H. P. Spaan and V. Mourai, "Pupil replication for exoplanet imaging," Astrophys. J. vol. 618, p. L165-L168 (2005).
34. F. H. P. Spaan and A. H. Greenaway, "Analysis of pupil replication," Astrophys. J. vol. 658, p. 1380–1385 (2007).
35. E. Serabyn and M. M. Colavita, "Fully symmetric nulling beam combiners," Applied Optics vol. 40, p. 1668-1671 (2001).
36. F. Cassaing, J.M. LeDuigou, J.P. Amans, M. Barillot, T. Buey, F. Hénault, K. Houairi, S. Jacquinod, P. Laporte, A. Marcotto, L. Pirson, J.M. Reess, B. Sorrente, G. Rousset, V. Coudé du Foresto and M. Ollivier, "Persee: a nulling demonstrator with real-time correction of external disturbances," Proceedings of the SPIE vol. 7013, n° 70131Z (2008).
37. L. Kaltenegger, M. Fridlund, "The Darwin mission: Search for extra-solar planets," Advances in Space Research vol. 36, p. 1114-1122 (2005).
38. TPF-I Science Working Group Report, JPL Publication 07-1, P.R. Lawson, O.P. Lay, K.J. Johnston and C.A. Beichman eds., Jet Propulsion Laboratory, California Institute of Technology, Pasadena, California (2007).
39. R. O. Gappinger, R. T. Diaz, A. Ksendzov, P. R. Lawson, O. P. Lay, K. M. Liewer, F. M. Loya, S. R. Martin, E. Serabyn, J. K. Wallace, "Experimental evaluation of achromatic phase shifters for mid-infrared starlight suppression," Applied Optics vol. 48, p. 868-880 (2009).
40. P. Haguenauer, E. Serabyn, "Deep nulling of laser light with a single-mode-fiber beam combiner," Applied Optics vol. 45, p. 2749-2754 (2006).

41. C. Buisset, X. Rejeaunier, Y. Rabbia, M. Barillot, "Stable deep nulling in polychromatic unpolarized light with multiaxial beam combination," Applied Optics vol. 46, p-7817-7822 (2007).
42. F. Hénault, "Fibered nulling telescope for extra-solar coronagraphy," accepted by Optics Letters.
43. C. G. Wyne, "Extending the bandwidth of speckle interferometry," Optical Engineering vol. 28, p. 21-25 (1979).
44. W. Benesch, J. Strong, "The Optical Image Transformer," JOSA vol. 41, p. 252-254 (1951).
45. F. Laurent, F. Hénault, P. Ferruit, E. Prieto, D. Robert, E. Renault, J.P. Dubois, R. Bacon, "CRAL activities on advanced image slicers: optical design, manufacturing, assembly, integration and testing," New Astronomy Reviews vol. 50, n° 4-5, p. 346-350 (2006).
46. F. Laurent, F. Hénault, E. Renault, R. Bacon, J.P. Dubois, "Design of an Integral Field Unit for MUSE, and results from prototyping," Publications of the Astronomical Society of the Pacific vol. 118, n° 849, p. 1564-1573 (2006).

## FIGURES CAPTIONS

1. Figure 1: Used reference frames on-sky, on the entrance and exit pupils, and in image plane.
2. Figure 2: Geometrical configurations of entrance and exit pupils.
3. Figure 3: Input and output optical layout of a generic multi-aperture, high angular resolution system.
4. Figure 4: Input and output sub-pupils configurations for the Fizeau-like interferometer (top) and hypertelescope (bottom).
5. Figure 5: Schematic layout of a hypertelescope.

6. Figure 6: Images formed by a hypertelescope for various baseline values B at $\lambda$ = 10 μm (original object shown on top left panel).
7. Figure 7: Conceptual optical layout of a super-resolving telescope.
8. Figure 8: Gray-scale map and slices along the U-V axes of the far-field fringe function F(M') projected on-sky by the SRT.
9. Figure 9: Illustration of the 4 x 4 mosaïcing procedure. The optical axis of the telescope is tilted by steps of 30 mas, scanning a 140 x 140 mas square angular area.
10. Figure 10: Numerical simulation of images formed by a super-resolving telescope. Top left, original object; top right, image at $\lambda$ = 10 μm formed by a 5-m telescope; bottom left, elementary image acquired by a 5-m SRT; bottom right, reconstructed image after a 4 × 4 mosaïcing and reconstruction process. Images sampling is 149 × 149.
11. Figure 11: Symmetric optical layout for co-axial recombination.
12. Figure 12: Images formed by an axial combining interferometer for various baseline values B at $\lambda$ = 10 μm (original object shown on top left panel. Images sampling is 439 × 439).
13. Figure 13: Interferometer equipped with nulling periscopes and a single-mode fiber centred on the optical axis (collecting and densifying optics are not shown).
14. Figure 14: Computed nulling maps for some typical cases. Top row: nulling Fizeau-like interferometer with two collecting telescopes. Second row: nulling SRT constituted of two exit recombining arms. Lower rows: axially combined interferometer with two, four and eight collecting telescopes (left: linear gray-scale; right: logarithmic gray-scale).
15.

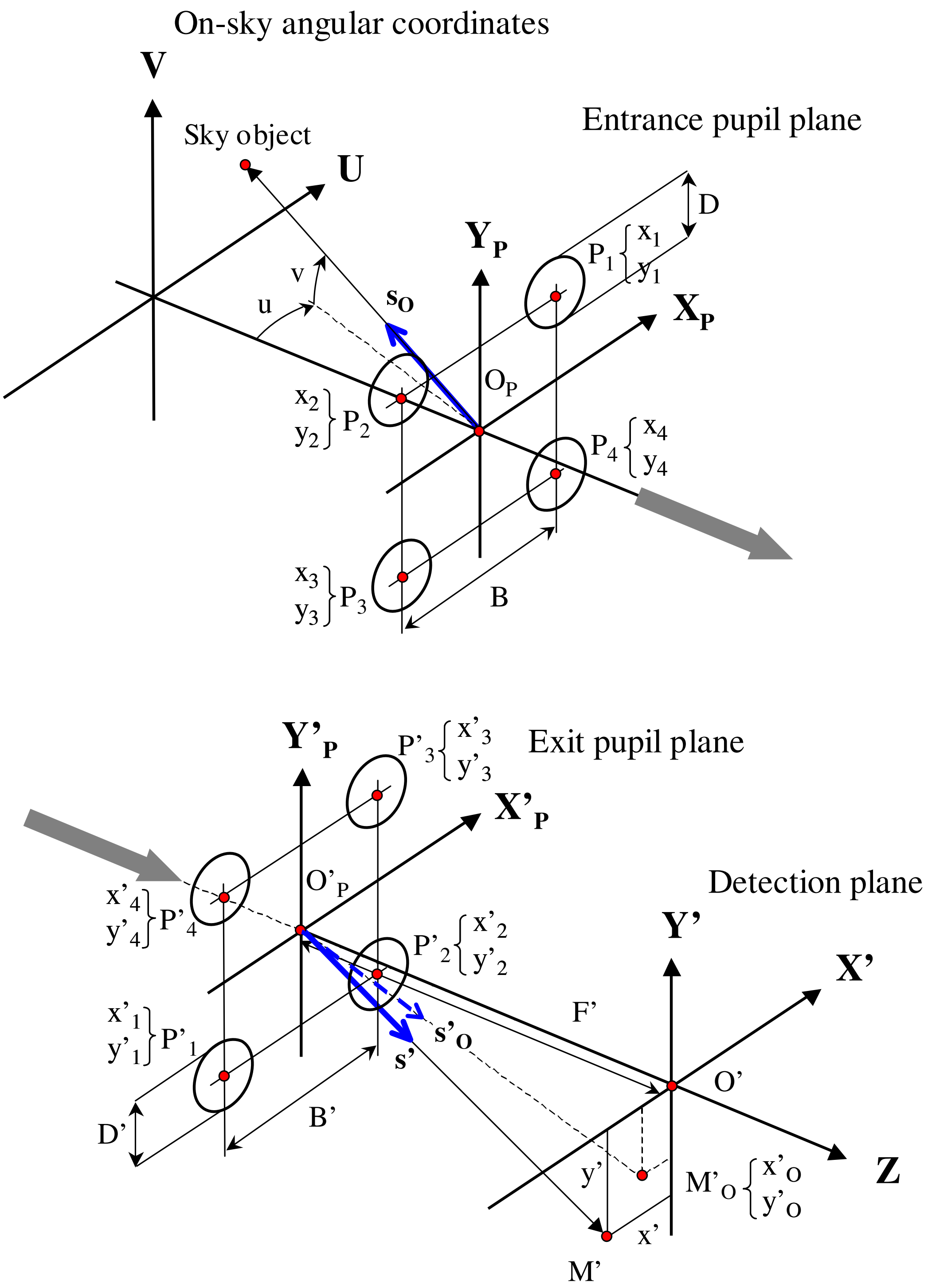

**Figure 1: Used reference frames on-sky, on the entrance and exit pupils, and in image plane.**

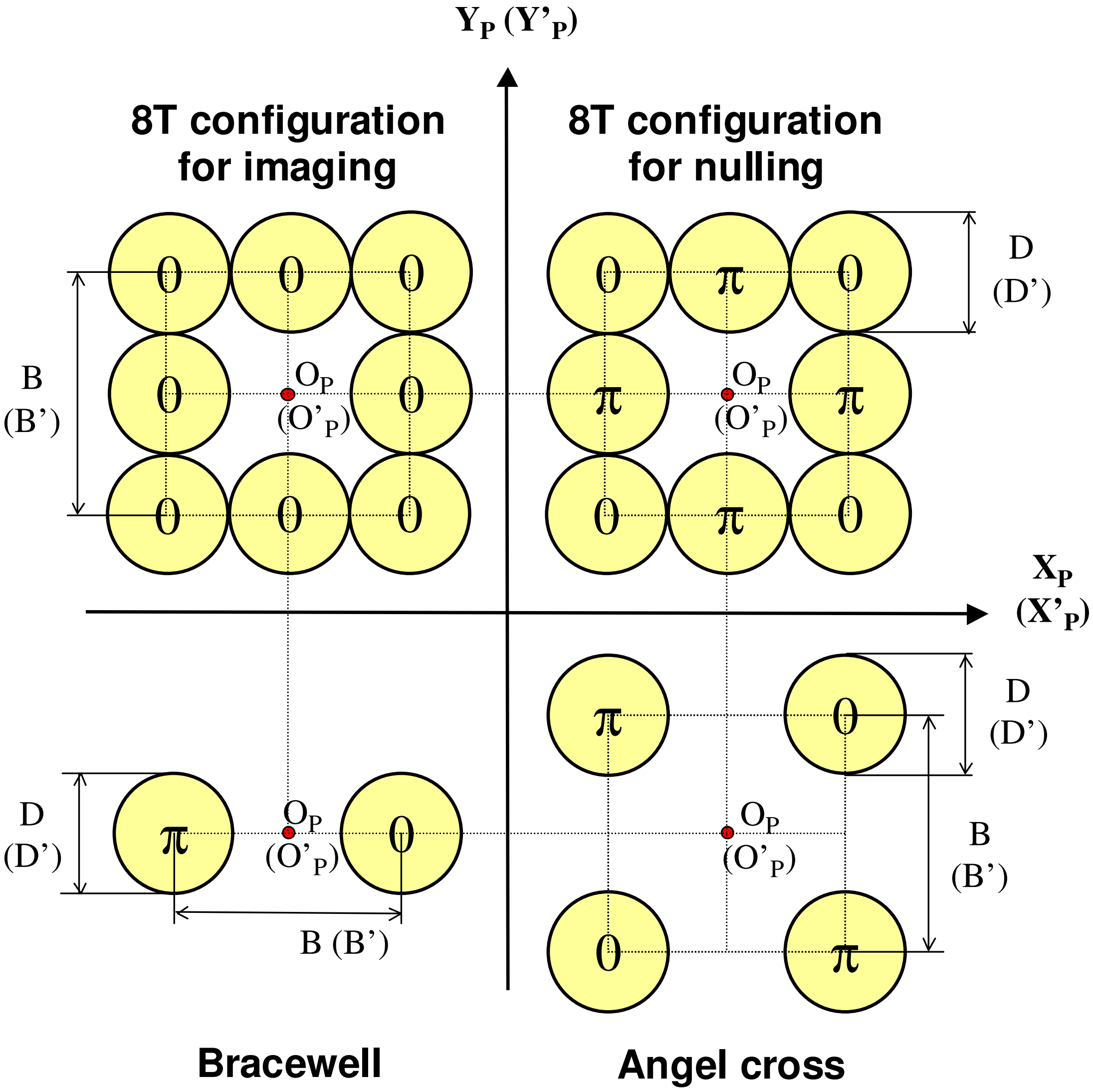

**Figure 2: Geometrical configurations of entrance and exit pupils.**

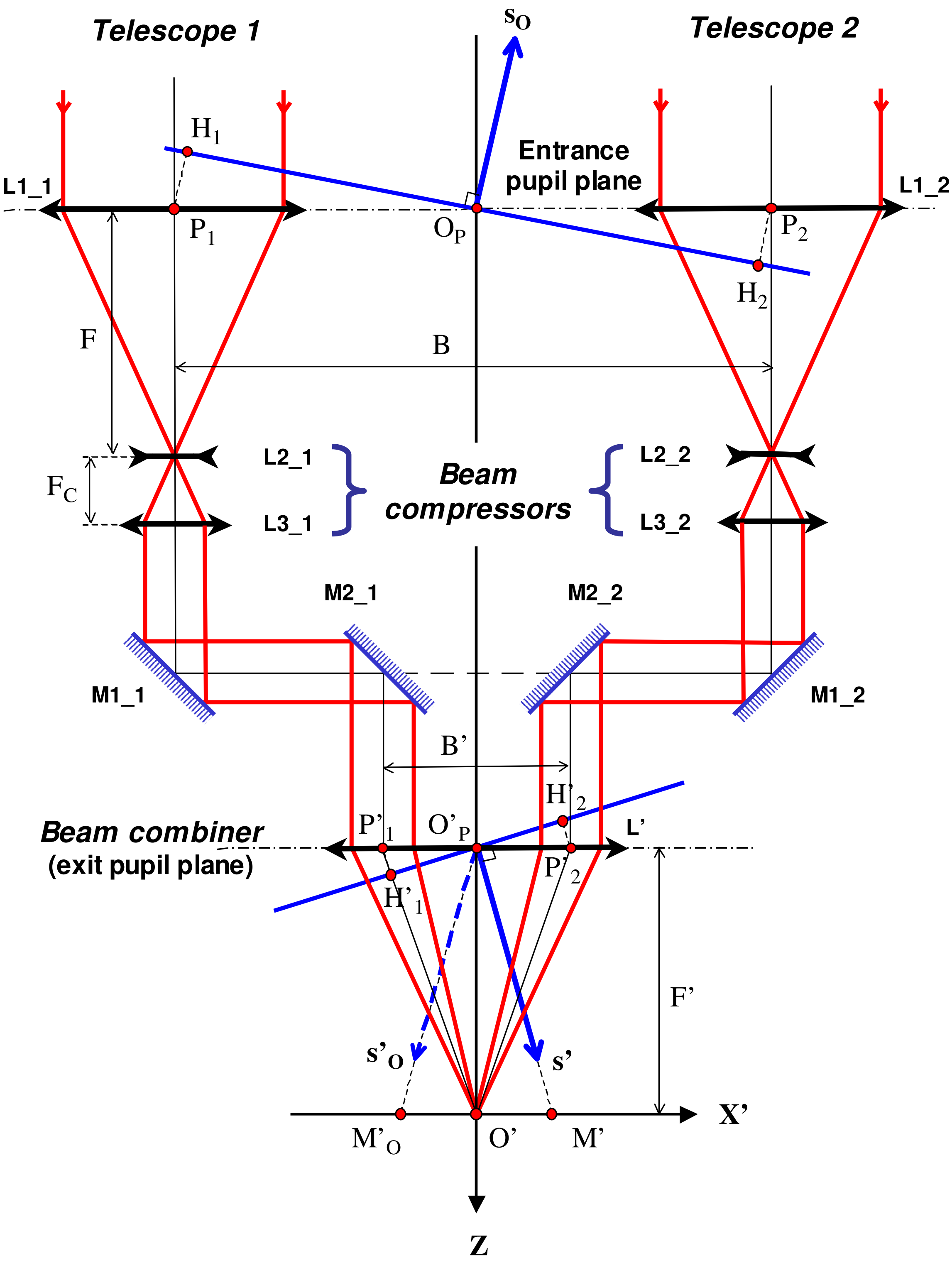

**Figure 3: Input and output optical layout of a generic multi-aperture, high angular resolution system.**

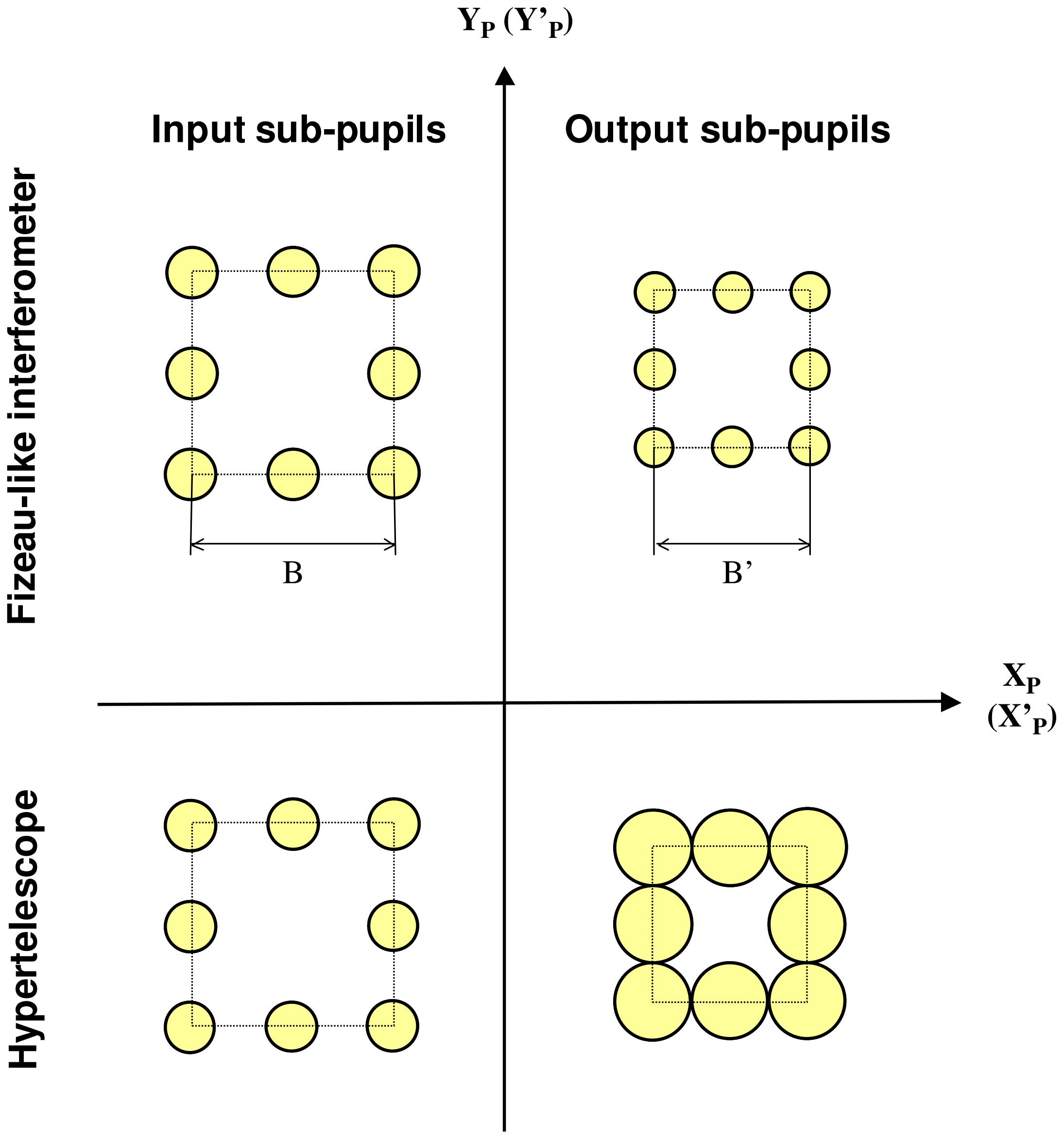

**Figure 4: Input and output sub-pupils configurations for the Fizeau-like interferometer (top) and hypertelescope (bottom).**

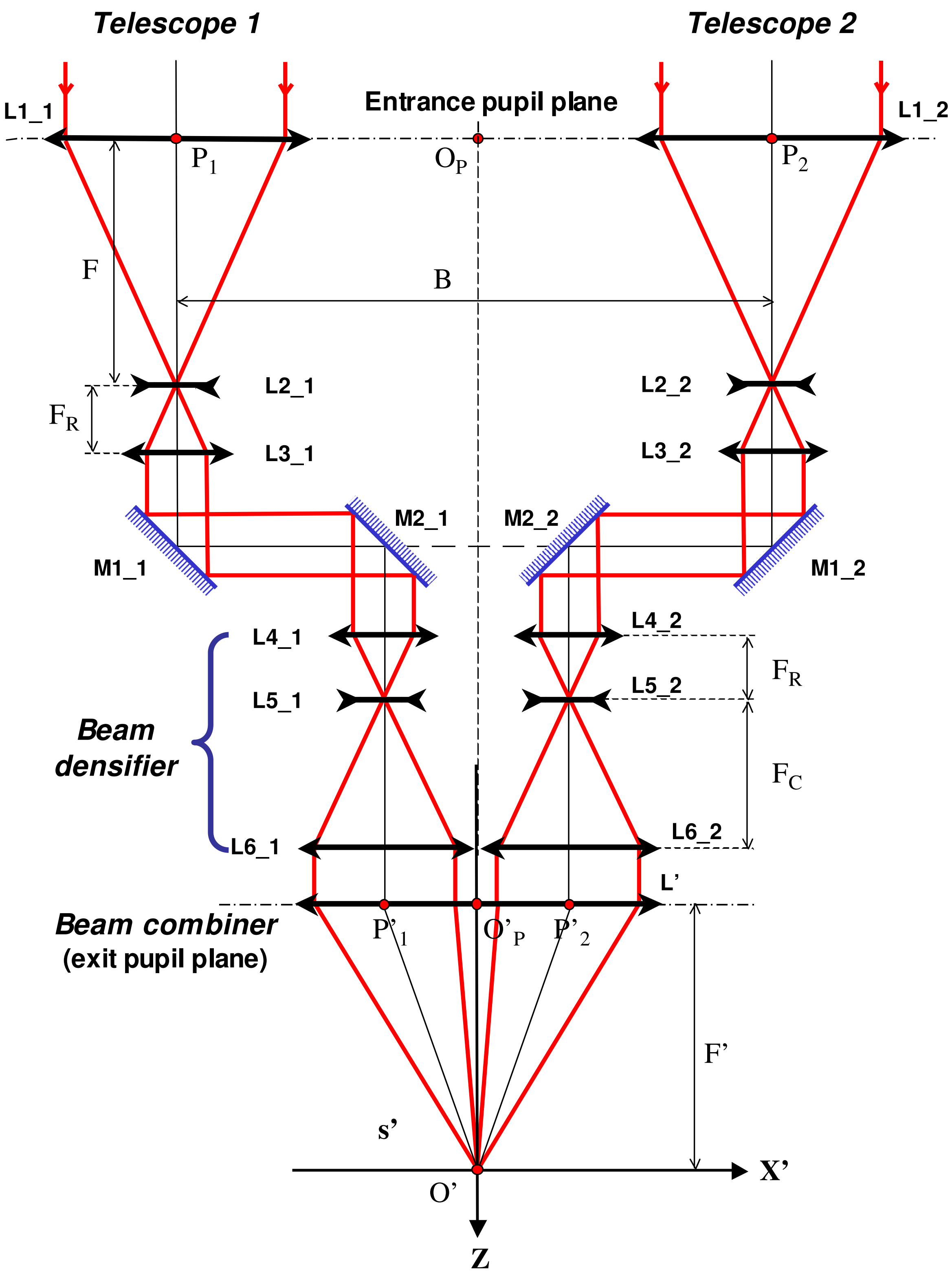

**Figure 5: Schematic layout of a hypertelescope.**

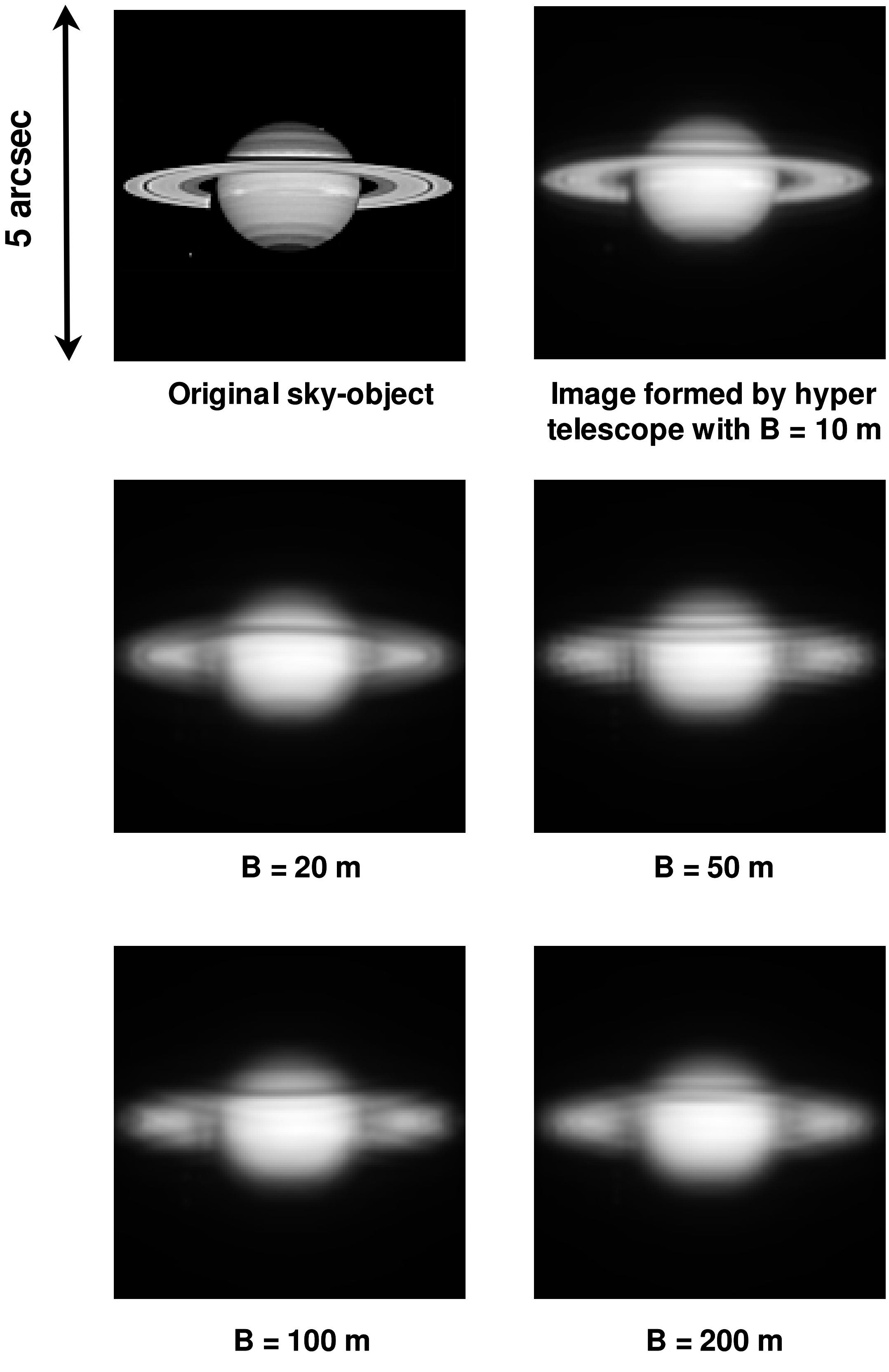

**Figure 6: Images formed by a hypertelescope for various baseline values B at $\lambda$ = 10 $\mu$m (original object shown on top left panel).**

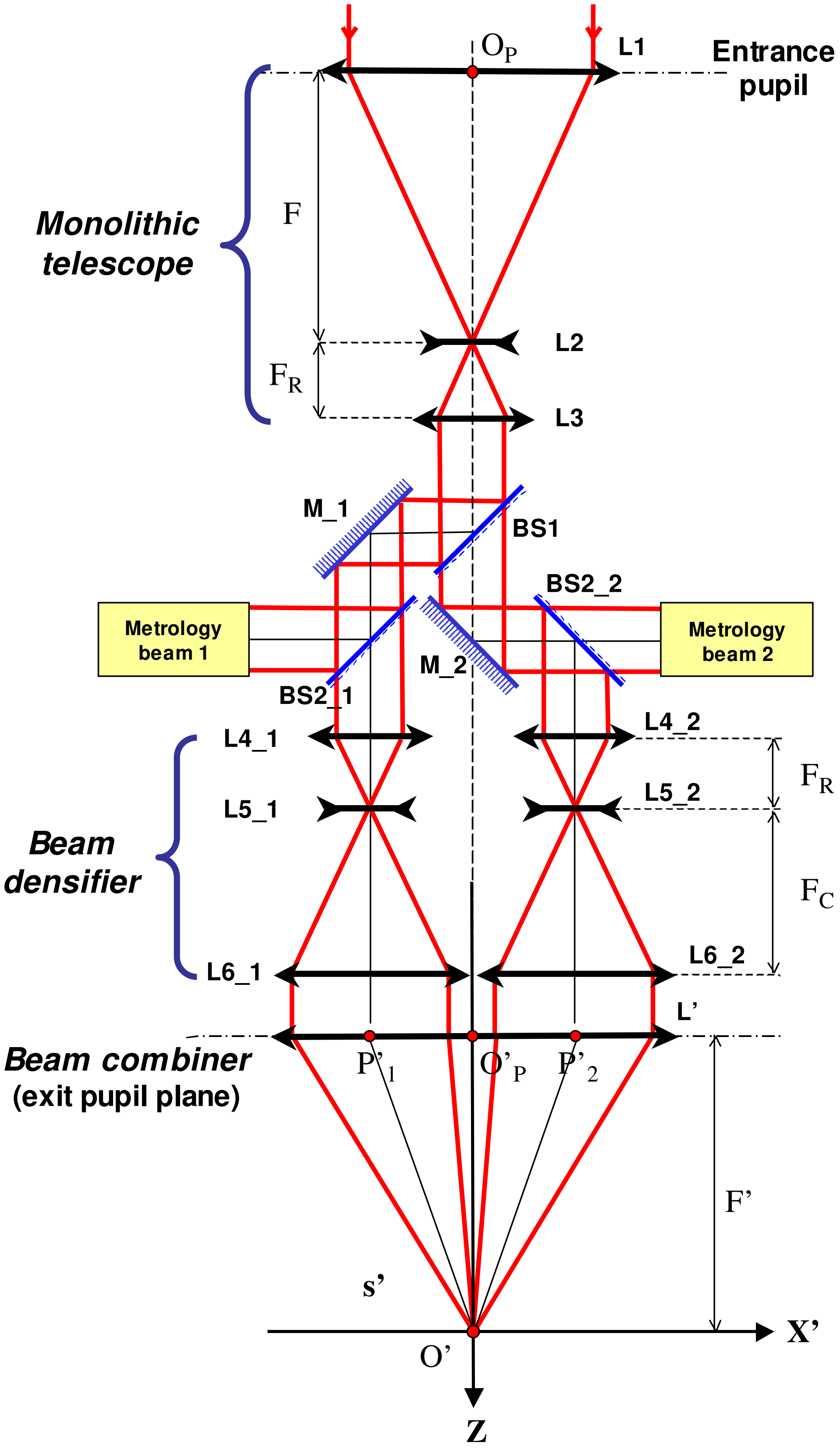

**Figure 7: Conceptual optical layout of a super-resolving telescope.**

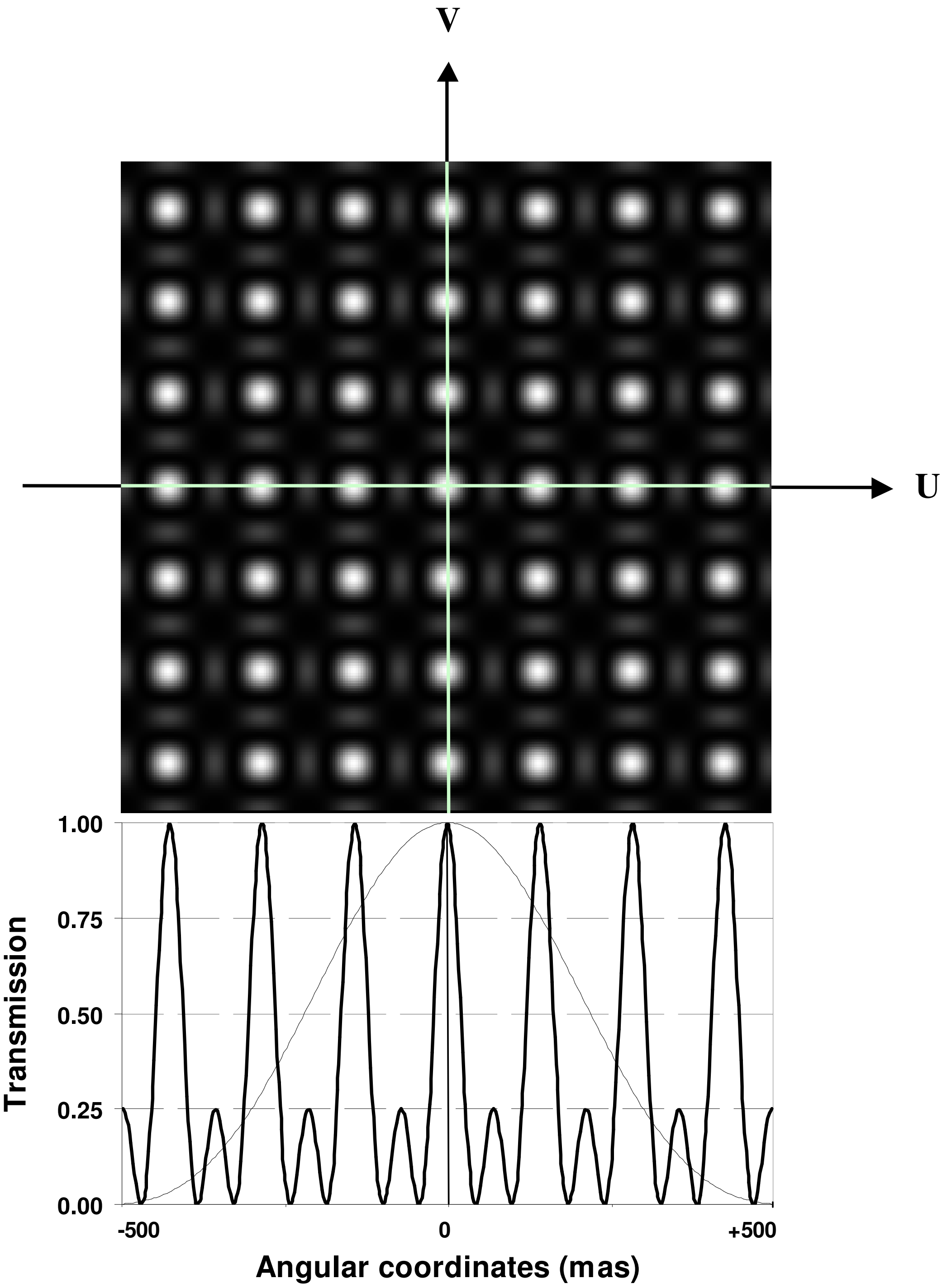

**Figure 8: Gray-scale map and slices along the U-V axes of the far-field fringe function F(M') projected on-sky by the SRT.**

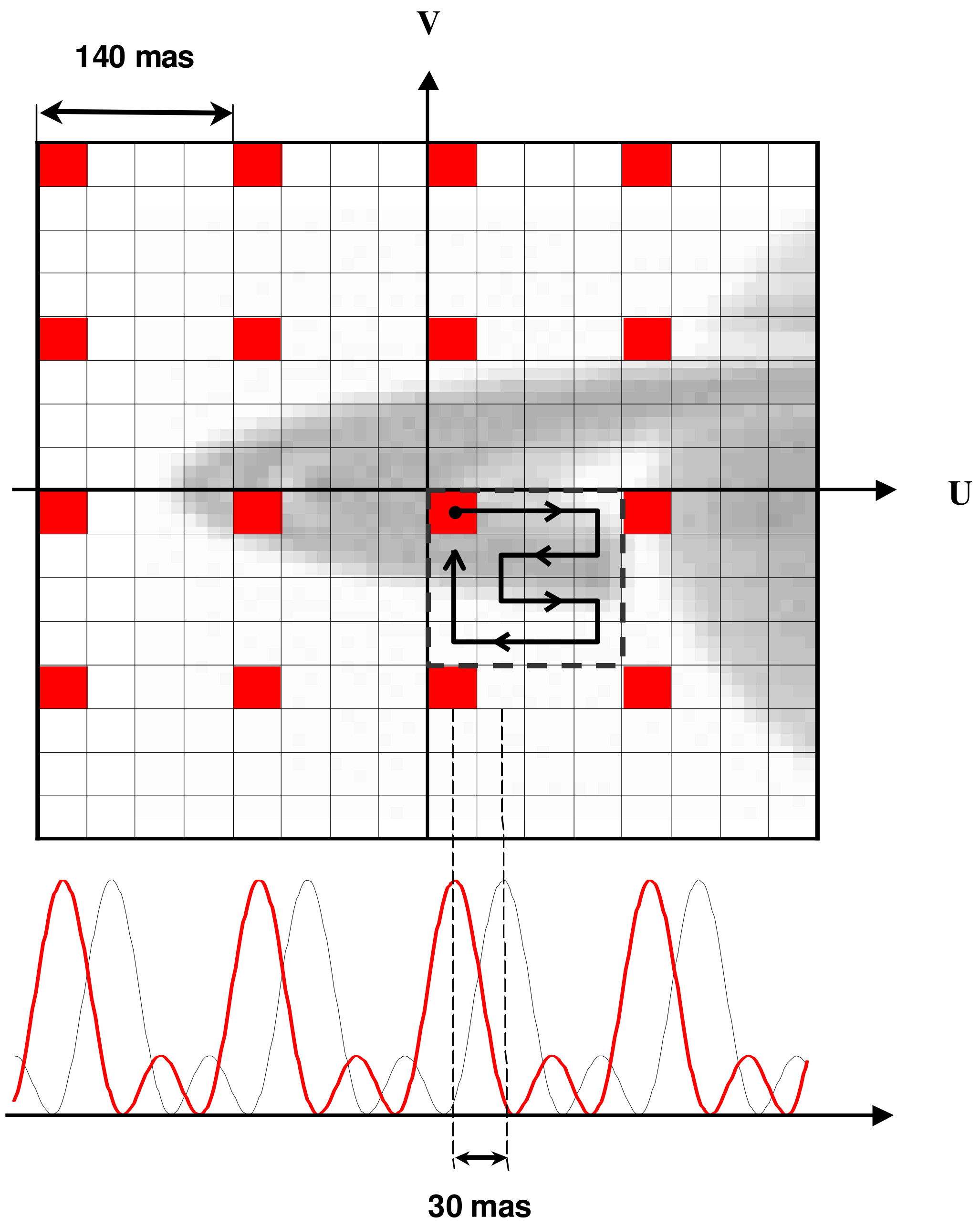

**Figure 9: Illustration of the 4 x 4 mosaïcking procedure. The optical axis of the telescope is tilted by steps of 30 mas, scanning a 140 x 140 mas square angular area.**

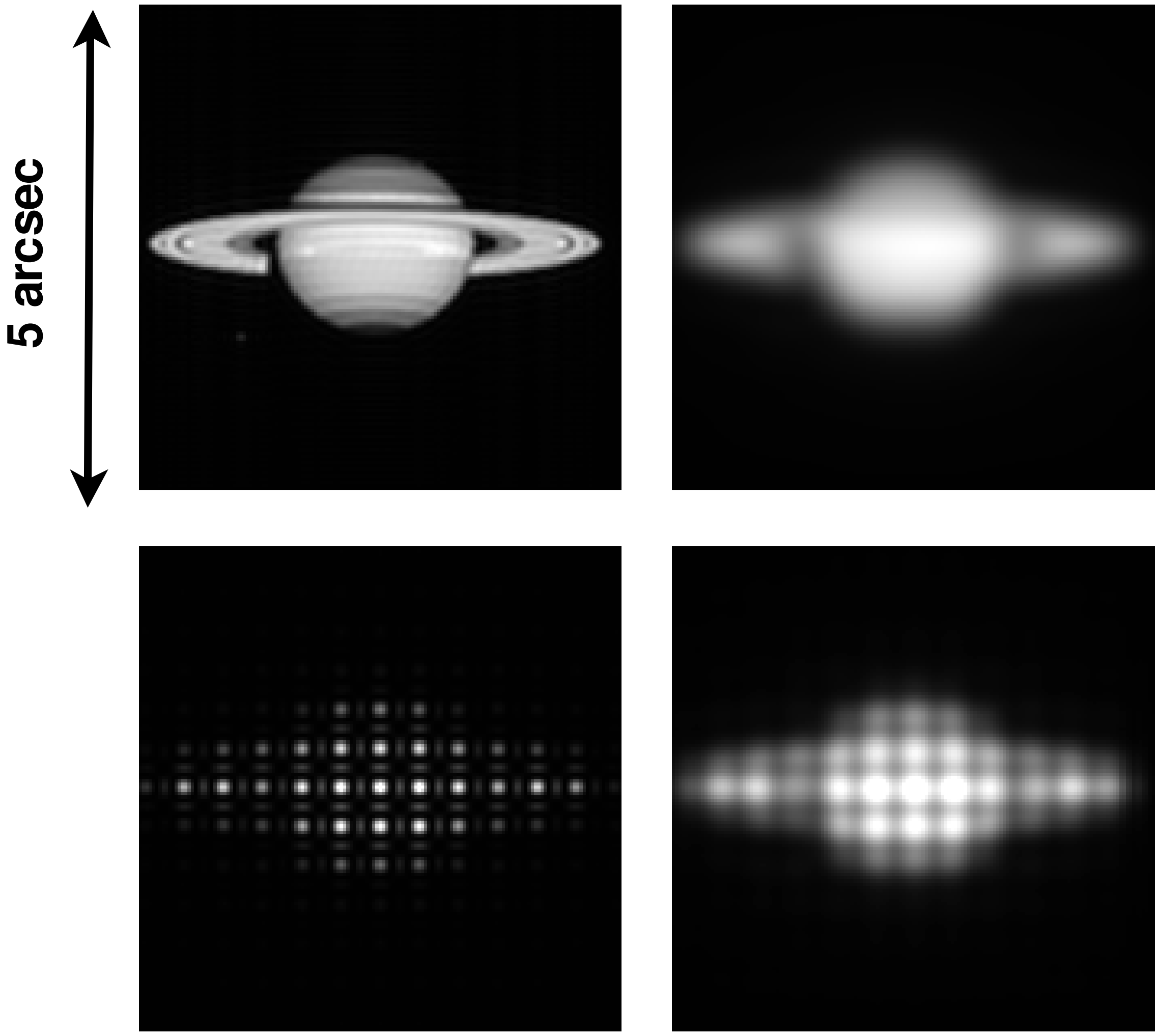

**Figure 10: Numerical simulation of images formed by a super-resolving telescope. Top left, original object; top right, image at $\lambda = 10$ µm formed by a 5-m telescope; bottom left, elementary image acquired by a 5-m SRT; bottom right, reconstructed image after a $4 \times 4$ mosaïcking and reconstruction process. Images sampling is $149 \times 149$.**

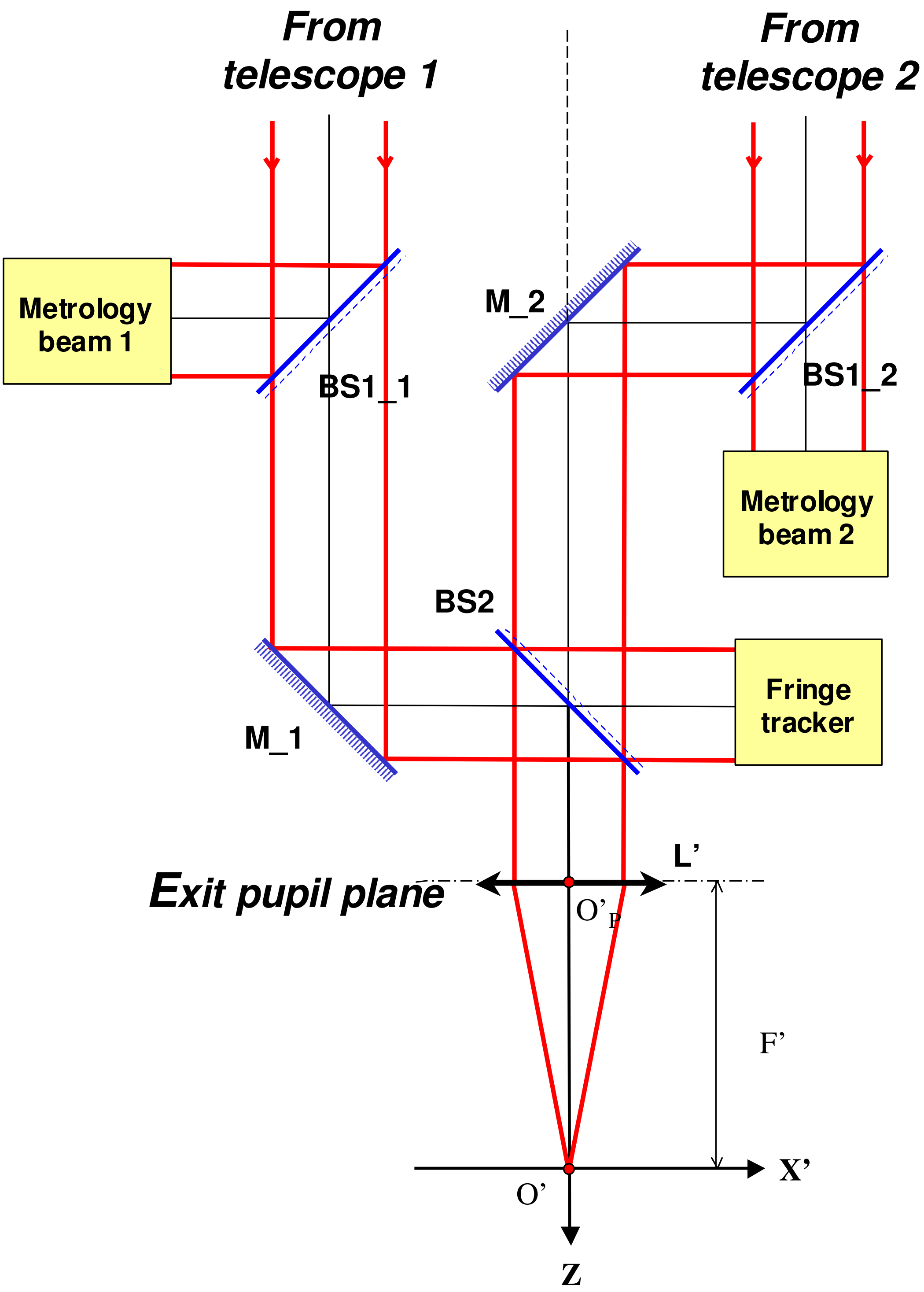

**Figure 11: Symmetric optical layout for co-axial recombination.**

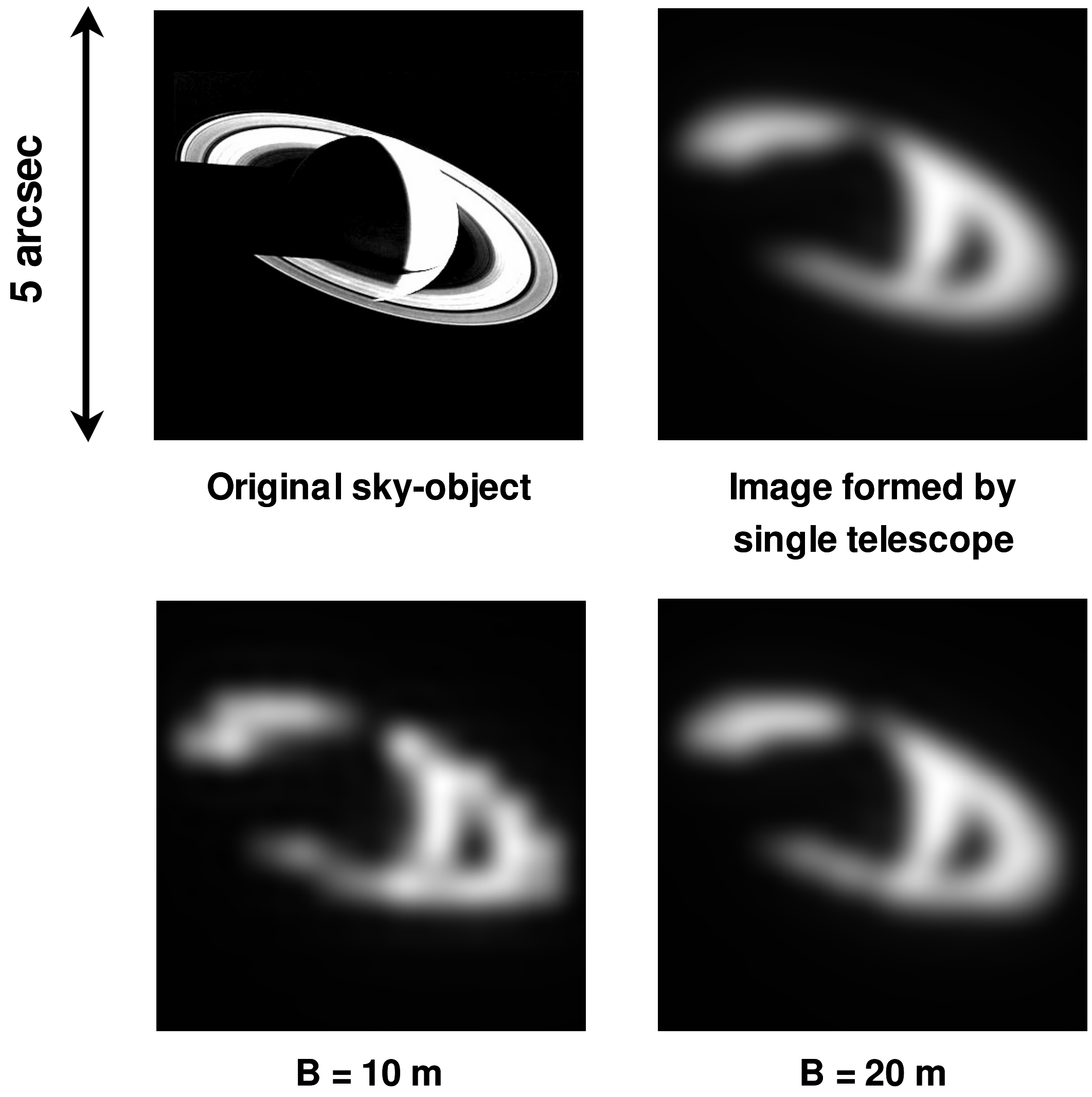

**Figure 12: Images formed by an axial combining interferometer for various baseline values B at $\lambda = 10$ µm (original object shown on top left panel. Images sampling is $439 \times 439$).**

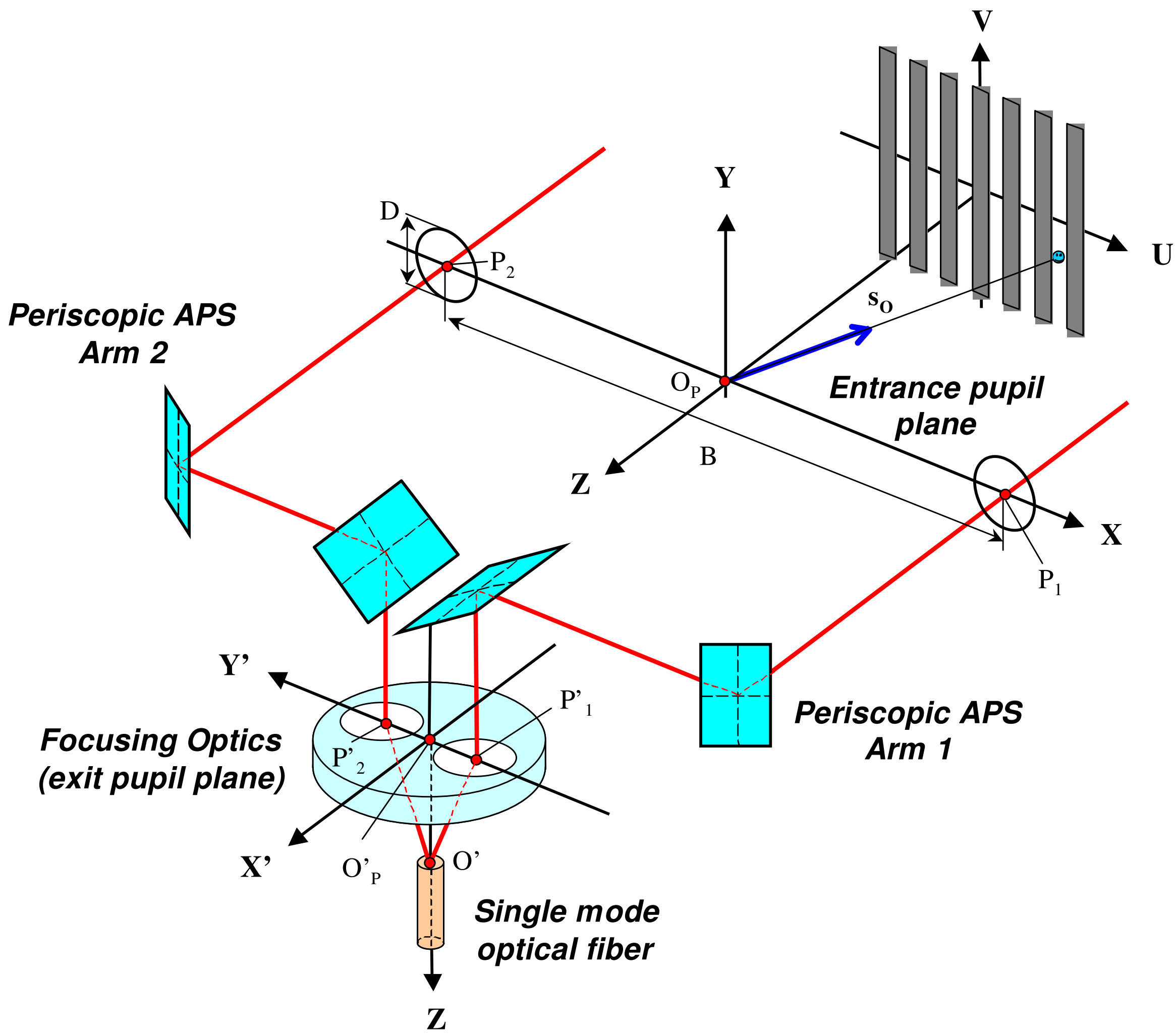

**Figure 13: Interferometer equipped with nulling periscopes and a single-mode fiber centred on the optical axis (collecting and densifying optics are not shown).**

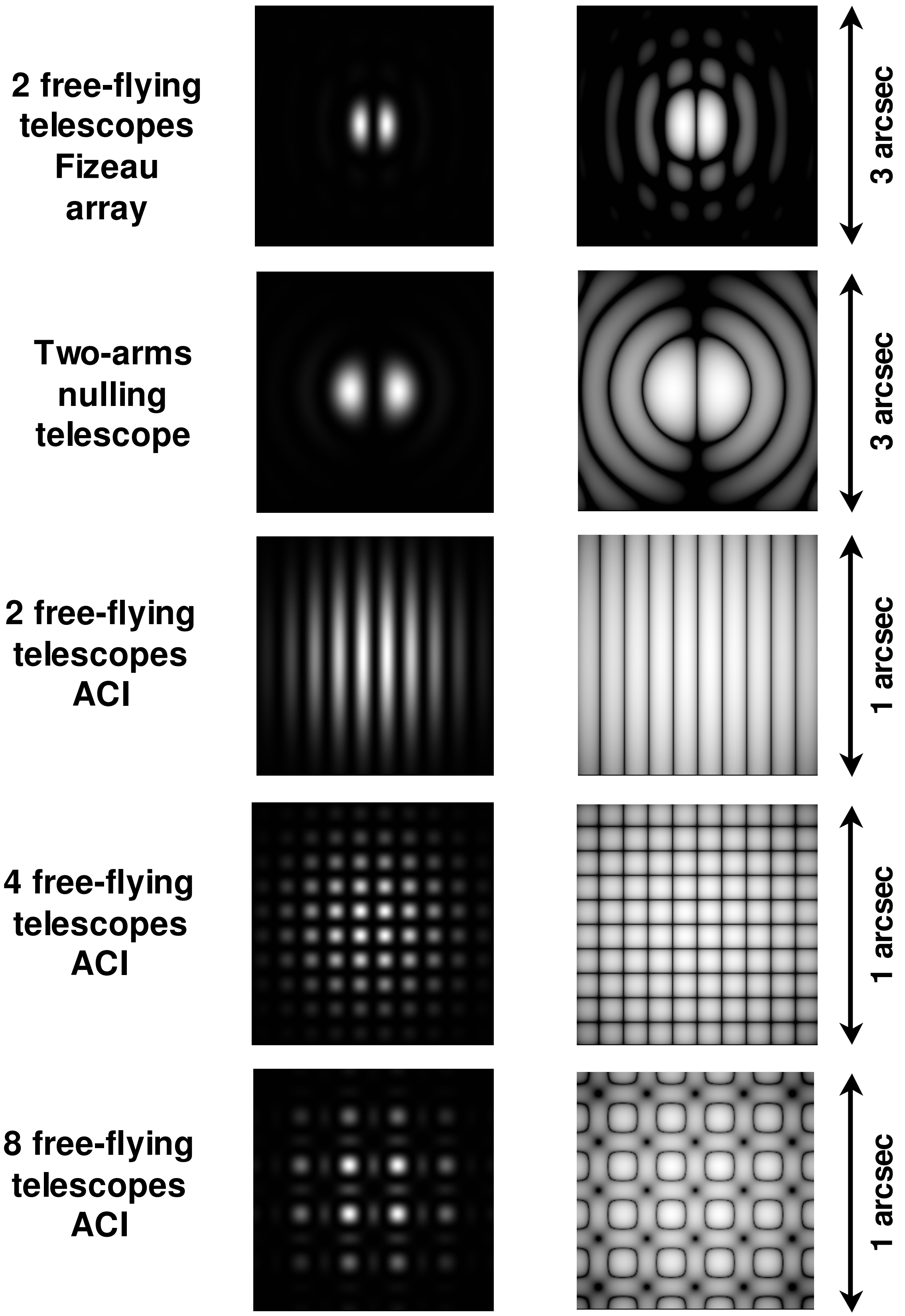

**Figure 14: Computed nulling maps for some typical cases. Top row: nulling Fizeau-like interferometer with two collecting telescopes. Second row: nulling SRT constituted of two exit recombining arms. Lower rows: axially combined interferometer with two, four and eight collecting telescopes (left: linear gray-scale; right: logarithmic gray-scale).**